\documentclass[3p, times, sort&compress]{elsarticle}
\usepackage{graphicx} %

\usepackage[utf8]{inputenc}

\usepackage{hyperref}
\usepackage{graphicx}
\graphicspath{{./Figures/}}
\usepackage{natbib}
\usepackage{amsmath}
\usepackage{amsfonts}
\usepackage{multirow}
\usepackage{threeparttable}
\usepackage{bm}
\usepackage{textcase}
\usepackage{pifont}
\usepackage{xcolor}

\usepackage{tikz}
\usetikzlibrary{calc,intersections}
\usetikzlibrary{spy}
\usepackage{pgfplots}
\usepackage{xcolor}
\usetikzlibrary{arrows}
\tikzstyle{line} = [draw, -latex', color=orange, line width=0.5mm]

\newcommand{\checked}{{\color{green}\ding{51}}}
\newcommand{\unchecked}{{\color{red}\ding{55}}}

\journal{Acta Astronautica}

\usepackage{bm}
\usepackage{subcaption}

\DeclareMathAlphabet{\mymathbb}{U}{bbold}{m}{n}

\newcommand{\SdR}[1]{\ensuremath{\mathcal{\MakeTextUppercase{#1}} = \left\{\MakeLowercase{#1},\,\MakeLowercase{\bm{#1}}_1,\,\MakeLowercase{\bm{#1}}_2,\,\MakeLowercase{\bm{#1}}_3\right\}}}
\newcommand{\sdr}[1]{\ensuremath{\mathbb{\MakeTextUppercase{#1}} = \left\{\MakeTextUppercase{#1},\,\MakeTextUppercase{\bm{#1}}_1,\,\MakeTextUppercase{\bm{#1}}_2\right\}}}

\title{RETINA: a hardware-in-the-loop optical facility with reduced optical aberrations}
\date{}

\author[1]{Paolo Panicucci}
\address[1]{Dipartimento di Scienze e Tecnologie Aerospaziali, Politecnico di Milano}
\author[1]{Fabio Ornati}
\author[1]{Francesco Topputo}

\begin{document}

	\begin{abstract}
		The increasing interest in spacecraft autonomy and the complex tasks to be accomplished by the spacecraft raise the need for a trustworthy approach to perform Verification \& Validation of Guidance, Navigation, and Control algorithms. In particular, in the context of autonomous navigation, vision-based navigation algorithms have established themselves as effective solutions to determine the spacecraft's state in orbit with low-cost and versatile sensors. Nevertheless, detailed testing must be performed on ground to understand the algorithm's robustness and performance on flight hardware. Given the impossibility of testing directly on orbit these algorithms, a dedicated simulation framework must be developed to emulate the orbital environment in a laboratory setup. This paper presents the design of a low-aberration optical facility called RETINA to perform this task. RETINA is designed to accommodate cameras with different characteristics (e.g., sensor size and focal length) while ensuring the correct stimulation of the camera detector. To design RETINA, a preliminary design is performed to identify the range of possible components to be used in the facility according to the facility requirements. Then, a detailed optical design is performed in Zemax OpticStudio to optimize the number and characteristics of the lenses composing the facility's optical systems. The final design is compared against the preliminary design to show the superiority of the optical performance achieved with this approach. This work presents also a calibration procedure to estimate the misalignment and the centering errors in the facility. These estimated parameters are used in a dedicated compensation algorithm, enabling the stimulation of the camera at tens of arcseconds of precision. Finally, two different applications are presented to show the versatility of RETINA in accommodating different cameras and in simulating different mission scenarios.
	\end{abstract}
	
	
	\maketitle

\section{Introduction}
Current space missions foresee complex and risky operations (e.g., in-orbit servicing, landing, or small body sampling) to achieve mission goals, prove technological demonstrations, and increase scientific return. These tasks cannot be performed in the context of present human-in-the-loop operations because the radio signal propagates to the spacecraft with finite velocity, implying delays that could lead to mission failure. Indeed, the spacecraft is required to be reactive and aware of changes in the external environment to fulfill mission objectives. Furthermore, the ground-based Deep Space Network is a crucial asset whose communication slots are limited in number and extremely costly. The increasing number of spacecraft raises questions about how to operate this increasing number of probes. One of the solutions is to limit the communication with the spacecraft and to let the probes operate autonomously without human supervision. These limitations underline the need to develop algorithms and methods to make spacecraft more autonomous. Therefore, the spacecraft must be able to sense the external environment, decode the information obtained from the sensors, and determine its state with respect to the external environment to enable complex operation and foster future exploration of the solar system. One of the possible solutions to perceive the external environment and determine the spacecraft within it is vision-based navigation (VBN). Cameras are usually preferred with respect to other complex sensors (such as LiDAR) because they do not deeply affect the spacecraft's mass, power, and size budgets. Moreover, image processing algorithms usually extract subpixel information, providing accurate and real-time measurements to navigation filters and perception algorithms. In addition, VBN is an approach compatible with all mission phases toward celestial bodies: cruise \cite{bhaskaran2000deep, andreis2022onboard, andreis2023autonomous}, approach \cite{panicucci2023shadow, panicucci2021autonomous, panicucci2023vision}, mid-range \cite{panicucci2023current,pugliatti2022data, franzese2019autonomous}, and close proximity \cite{piccolo2022simultaneous, norman2022autonomous, maass2020crater}.\newline
A crucial step to be addressed when developing a complex VBN algorithm is Validation and Verification (V\&V). First, it is hard to perform tests on orbit, even when the algorithm has a high Technology Readiness Level (TRL). The lack of halfway testing on the orbiting platform implies high risks to the final product, and it increases costs if issues arise at late-stage development. Because of this reason, different approaches must be foreseen to assess the correct deployment, integration, and performance of vision-based systems. Current approaches rely on the generation of realistic datasets on ground to assess the algorithm performance and the computation burden on hardware components. In the context of complex spacecraft operations, such as on-orbit servicing and deep-space applications, it is complex to gather these datasets from previous missions. Therefore, V\&V of the VBN chain is assessed incrementally, by tackling a single feature individually in different validation environments to identify criticalities at an early stage of the development. First, numerical simulations are performed by simulating cameras with dedicated rendering engines, such as ESA's PANGU \cite{rowell2012pangu} or Airbus Defence \& Space's SurRender \cite{lebreton2021image}. Second, processor-in-the-loop (PIL) simulations are performed to assess the algorithm execution and latency on representative hardware. Third, camera-in-the-loop (CIL) simulations are put in place to understand the image processing performance in the presence of the camera optical system which can degrade the VBN estimates due to distortions and aberrations. To perform these tests in a laboratory environment is not an easy task, as the camera is a geometrical and radiometric sensor, implying that it must observe the correct geometrical scene with a light signal consistent with what would be observed in orbit. To do so, CIL testing is performed in two different types of facilities aiming at generating realistic image datasets: optical and robotic facilities. On the one hand, optical facilities are test-benches where the camera is stimulated from a screen via a lens system. These facilities have the drawback of partially relying on image generation, but their development and maintenance costs are limited. On the other hand, robotic facilities are composed of one or more robotic components, generally arms, moving the sensor, and the target mock-up to simulate the needed observational geometry. These facilities are very realistic in terms of image content as light physically propagates from the target to the detector, but their cost is high and the reproducible observational geometries are limited due to displacement limitations of the bulky robotic infrastructure. To overcome both limitations, they are usually used in a complementary manner to increase the validation test realism incrementally. Indeed, optical facilities are usually used to preliminary assess the algorithm robustness before moving to more complex and costly robotic facility testing. \newline
This work aims to present the design and the performance of the RETINA optical facility. RETINA (Realistic Experimental faciliTy for vIsion-based NAvigation) is designed to accommodate cameras with different optical head and detector characteristics by ensuring light collimation with a reduced level of chromatic and achromatic aberrations. RETINA can be used with different cameras thanks to its two-lens-assembly design ensuring that the screen magnification is consistent with the camera field of view (FoV). Moreover, the lens assemblies are designed with commercial-of-the-shelf (COTS) components optimized to reduce the chromatic and achromatic aberration due to the multi-lens design. Moreover, a dedicated geometrical calibration algorithm is designed to achieve subpixel precision and accuracy to compensate for errors due to RETINA distortions and misalignment. These hardware and software components ensure that the camera mounted in RETINA can observe the scene projected on the screen to correctly assess the VBN chain performance with the camera in the simulation loop.\newline
%The rest of the paper is organized as follows. {\color{red} Section}

\section{Notation}\label{sec:notation}
In this document, the following notation is used:
\begin{itemize}
	\item 3D and 2D vectors are denoted respectively with lower and upper case bold text, such as $\bm{r}$ and $\bm{R}$.
	\item Matrices are in plain text in brackets, such as $\left[A\right]$.
	\item Vector initialization is performed with parenthesis, such as $\bm{b} = \left(\bm{a}^T\; \bm{a}^T\right)^T$.
	\item \SdR{a} is the 3D reference frame centered in the 3D point $a$ with axes $\bm{a}_1$, $\bm{a}_2$, and $\bm{a}_3$. All the reference frames are right-handed and orthonormal.
	\item The rotation matrix from $\mathcal{S}$ to $\mathcal{C}$ is $\left[CS\right]$. All rotations have a passive function.
	\item \sdr{a} is the a 2D reference frame centered in the 2D point $A$ with axes $\bm{A}_1$ and $\bm{A}_2$ which are orthonormal. 
	\item The 3D vector $\bm{r}$ in homogeneous form is labeled ${}_{h}\bm{r}$ and the 2D vector $\bm{R}$ in  homogeneous form is labeled ${}_{h}\bm{R}$.
	\item The projection of the 3D vector $\bm{r}$ on the 2D image is labeled $\bm{R}$.
\end{itemize}

\section{Previous Literature on Hardware-In-the-Loop Optical Facility}\label{sec:SOTA}
Optical facilities were designed in the past to assess the star tracker performance on ground \cite{rufino2002laboratory, rufino2013real, filipe2017miniaturized, samaan2011star, nardino2019ministar}. To the authors' knowledge, the first work in the literature reporting the development and the use of an optical facility is \citet{rufino2001stellar}. \citet{rufino2001stellar} presents the design of an optical facility composed of a cathode ray tube display stimulating a star tracker through a collimator. This work analyzes the geometrical and radiometrical requirements the component must fulfill to enable the correct star tracker stimulation. Moreover, the analytical performance attainable in this category of facilities is initially reported in \citet{rufino2001stellar} and investigated in more detail in \citet{rufino2002laboratory}. A similar design with more recent technology is reported in \citet{rufino2013real} where the cathode ray tube display is substituted with an LCD screen. In these works, the geometrical calibration of the facility is obtained by estimating the misalignment and the optics distortion with neural network \cite{rufino2000effective}, while the radiometric calibration is obtained by exploiting the knowledge of the screen mapping from illuminance to digital number (DN) thanks to calibration curves provided by the manufacturer \cite{rufino2002laboratory, rufino2013real, rufino2001stellar}.\newline
A similar design is presented in \citet{filipe2017miniaturized} to test the ADCS subsystem of the MarCO and ASTERIA CubeSats. To perform this test, the facility had to be accommodated on an air-bearing system to perform closed-loop ADCS simulations. Therefore, \citet{filipe2017miniaturized} focuses on the miniaturization of the facility thanks to an OLED smartphone display mounted at the end of a rigid bar. This enables stimulating the camera mounted on the air-bearing system with collimated light while reaction wheels control the platform. The facility is geometrically calibrated, but not radiometrically. Indeed, the work details a procedure to tackle the problem of radiometric calibration, but the detailed method does not achieve satisfactory results as the assumption of a linear map between DN and illuminance does not hold.\newline
For larger star trackers, Jena-Optronik's Optical Sky field sImulator (OSI)  \cite{samaan2011star} and MINISTAR \cite{nardino2019ministar} were designed to test hardware after integration of the components on the spacecraft. The test benches are geometrically calibrated to correctly stimulate the camera with a miniaturized OLED microdisplay through a collimator. OSI is radiometrically calibrated by the manufacturer, while no information is available about the radiometric calibration procedure for MINISTAR.\newline
Recent interest in autonomous vision-based navigation (VBN) and the need for their validation have motivated the application of optical facilities to vision-based systems \cite{roessler2014optical, beierle2017design, beierle2017high, panicucci2022tinyv3rse}.
\citet{roessler2014optical} presents an optical test bench similar to the one outlined in  \citet{rufino2013real}, but it is used in the context of spacecraft rendevous VBN algorithm validation. The facility is geometrically calibrated by compensating the facility-induced errors with an estimated homography. Stanford's SLAB also developed its facility to test the developed vision-based algorithms \cite{beierle2017design, beierle2017high}. \citeauthor{beierle2017design} \cite{beierle2017design, beierle2017high} present a radiometrically and geometrically calibrated design to test autonomous formation flight navigation. The design is a collimated single-lens system where an OLED microdisplay is used to stimulate the camera. The radiometric calibration is performed by measuring the mapping between the screen DN and the displayed celestial object irradiance via a powermeter. Moreover, the geometrical calibration is performed by fitting a high-order polynomial to compensate for both misalignments and distortions induced by the facility. The obtained facility can display celestial objects with a wide range of irradiance with a projection error of less than tens of arcseconds. An improved design is presented in \citet{beierle2019variable} where a double-lens system is exploited to obtain variable magnification of the facility. This variable magnification implies that cameras with different FoVs can be tested with the same facility simply by regulating the relative distance between the lenses. These improvements come at the cost of a more complex optical design inducing high distortions and aberrations. To the authors' understanding, the facility design includes two cemented achromatic doublets to magnify the screen image to fit the screen height with the camera FoV. It is worth mentioning that this facility has been recently used to validate on-ground autonomous vision-based navigation for formation flight application \cite{kruger2023starling}.\newline
A single-lens collimated design is also presented in \citet{panicucci2022tinyv3rse} where the facility is designed to test space exploration scenarios (e.g. deep-space and asteroid navigation). In this work, geometrical calibration is addressed by presenting two different approaches to compensate for facility-induced errors: upstream and downstream compensations. In the former, errors are compensated before projecting the image on the screen, while the latter approach compensates errors after the camera acquisitions induced by the hardware-in-the-loop (HIL) testing. In both cases, the compensation is achieved by using a high-order polynomial to fit the facility-induced distortions. The radiometric calibration is not presented in \citet{panicucci2022tinyv3rse} even though preliminary results are presented in \citet{andreis2023towards}.\newline
In this context, this work aims to present RETINA, a HIL optical facility used to test image processing and vision-based navigation algorithms. RETINA differs from the previous design because its optical design is optimized to show superior optical performance thanks to optimized lens systems.\newline
A summary of the state-of-the-art optical facilities is reported in Table~\ref{tab:facilities} for the sake of completeness. 
\begin{table}[h]
	\centering 
	\caption{Summary of the optical facilities developed in the past for attitude and navigation testing.}
	\label{tab:facilities}
	\begin{threeparttable}
		\begin{tabular}{c|cccc}
			& Radiometrically & Geometrically & Reduced optical & Variable \\
			& calibrated & calibrated & aberrations & magnification\\
			\hline
			UniNa Facilities \cite{rufino2001stellar, rufino2002laboratory, rufino2013real} & \checked & \checked & NA\tnote{1} & \unchecked \\
			JPL Facilities \cite{filipe2017miniaturized} & \unchecked & \checked & NA\tnote{1} & \unchecked \\
			OSI \cite{samaan2011star} & \checked & \checked & NA\tnote{1} & \unchecked \\
			MINISTAR \cite{nardino2019ministar} & \unchecked & \checked & NA\tnote{1} & \unchecked \\
			UniDenmark Facilities \cite{roessler2014optical} & \unchecked & \checked & NA\tnote{1} & \unchecked \\
			TinyV3RSE \cite{panicucci2022improvements, panicucci2022tinyv3rse, pugliatti2022tinyv3rse} & \checked  \tnote{2} & \checked & NA\tnote{1} & \unchecked  \\
			Stanford Single-lens Facility \cite{beierle2017design, beierle2017high} & \checked & \checked & NA\tnote{1} & \unchecked \\
			Stanford Two-lens Facility \cite{beierle2019variable} & \checked & \checked & \unchecked & \checked \tnote{3} \\
			RETINA & \checked \tnote{2} &  \checked & \checked & \checked \tnote{4} \\
		\end{tabular}
		\begin{tablenotes}
			\item[1] Not applicable as single lens design. 
			\item[2] RETINA and TinyV3RSE radiometric calibrations are not presented in this work. Preliminary results are available in \citet{andreis2023towards, ornati2024retina}. 
			\item[3] Variable magnification can be also achieved during the simulation thanks to motorized optical stages.
			\item[4] Variable magnification can be only achieved before the calibration to match the camera mounted in the facility.
		\end{tablenotes}
	\end{threeparttable}
\end{table}

\section{Overview of RETINA}\label{sec:overview}
\begin{figure}[h]
	\centering
	\begin{tikzpicture}
		\node[] at (0,0){\includegraphics[ width=0.9\textwidth]{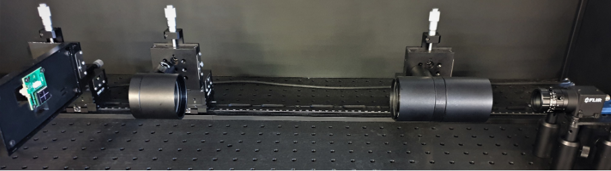}};
		\node[] at (7.5,2){\includegraphics[ width=0.15\textwidth]{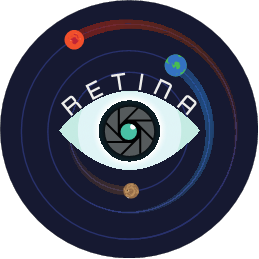}};
		\node[] at (6.5,-1) (camera){}; 
		\node[] at (3,-1) (collimator){}; 		\node[] at (-3.5,-1) (relay){}; 
		\node[] at (-6.5,-0.5) (screen){}; 
		\node[align=center,text width=3cm , color = black] at (+4.5,-3.75) (cameraText){Camera}; 
		\node[align=center,color = black] at (2,-3.15) (collimatorText){Collimator Lens System}; 
		\node[align=center,color = black] at (-2,-3.15) (relayText){Relay Lens System}; 
		\node[color = black] at (-5,-3.75) (screenText){OLED Microdisplay}; 
		\path[line] (cameraText)--(camera);
		\path[line] (collimatorText)--(collimator);		
		\path[line] (relayText)--(relay);
		\path[line] (screenText)--(screen);
	\end{tikzpicture}
	\caption{A picture of RETINA with its components underlined.}
	\label{fig:RETINA}
\end{figure}
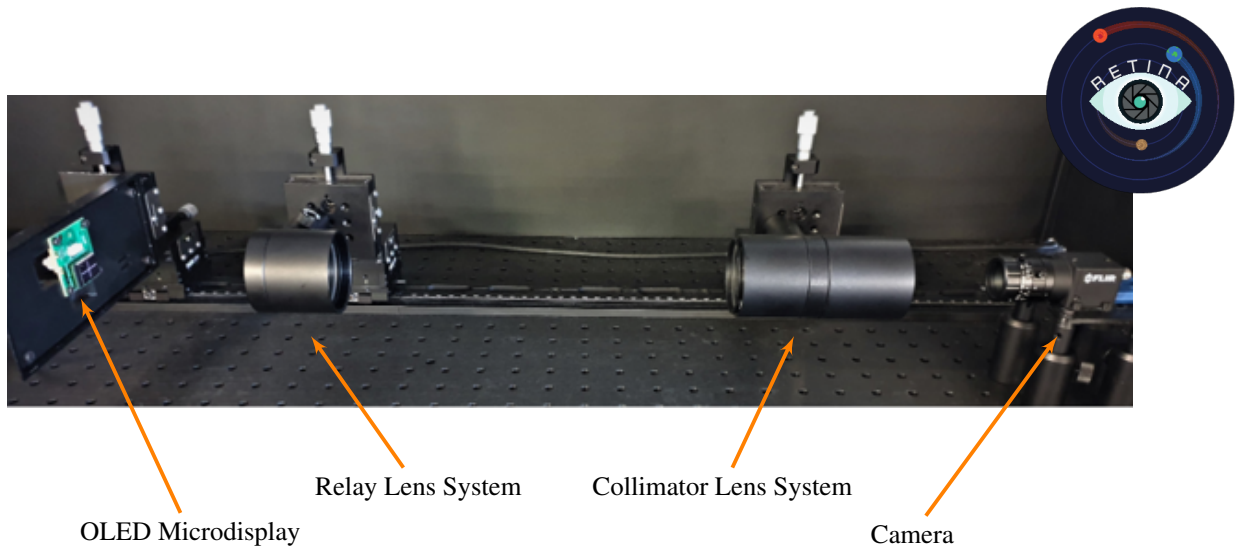

RETINA is an optical facility designed to test and validate vision-based navigation algorithms in hardware-in-the-loop simulations to emulate the sensor condition in orbit. When mounted in RETINA, spaceborne cameras are stimulated as they were in orbit with limited distortion and aberration effect due to optical components of the facility and hardware misalignment. To achieve so, RETINA is designed with optimized hardware to correctly stimulate cameras with different FoV (i.e., sensor size and focal lengths). The high-level requirements on the facility state that the facility shall be compliant with FoV between 4 and 22 degrees. This range is selected as it is compatible with cameras used in DART Lab projects (e.g., LUMIO \cite{panicucci2024vision} and Milani \cite{pugliatti2023vision}). Moreover, this range is compliant with the majority of space-graded hardware used for vision-based navigation, attitude determination, and scientific observations. RETINA is composed of several components that are mounted on a 60 cm $\times$ 120 cm optical breadboard and enclosed in a closed black box to limit external light pollution. In detail, the components are:
\begin{enumerate}
	\item A controllable OLED microdisplay.
	\item A relay lens system that magnifies the screen to perfectly fit the camera FoV.
	\item A collimator lens system that collimates the magnified screen to emulate space observations.
	\item A camera that is stimulated by the screen through the two lens systems.
\end{enumerate}
A picture of RETINA is reported in Figure \ref{fig:RETINA} showing the different components of the facility.
\newline 
The screen is selected to have a small size (i.e., 18.7 mm $\times$ 11.75 mm) to ensure a compact facility and a pixel pitch of 9.6 $\mu$m to enable the display of small objects. The resolution is  1920 $\times$ 1200 pixels ensuring to display small features and unresolved celestial objects in the camera field of view. Moreover, it relies on OLED technology, implying an extreme dynamic range between the low and high-illuminated pixels and the absence of backlight leakage for inactive pixels. This ensures have high contrast between the illuminated celestial objects and the dark background. Moreover, it is worth noting that a high illumination dynamic range is beneficial to display a space scene as celestial objects usually exhibit large variations in magnitude between faint stars and bright resolved objects. The screen is attached to a custom interface which is mounted to an opto-mechanical stage that can be regulated with millimeter precision. This stage, as the other ones in the facility, is mounted on a translational rail that enables the relative displacement of the components.\newline
Two lens systems are present in the facility, having respectively a very precise scope. In this section, a qualitative description of these lens systems is provided. A more detailed analysis of the lens systems preliminary design is reported in Section \ref{sec:twolensmodel}, while the detailed optical design is reported in Section \ref{sec:lowaberrationsdesign}.\newline
The collimating lens system aims at collimating the light that stimulates the camera to emulate space observations. Indeed, observed celestial objects are usually distant with respect to the camera, leading to the possibility of considering the light ray parallel when entering the camera pupil \cite{beierle2019variable, rufino2002laboratory}. The relay lens system is included to magnify the screen to simulate the presence of a smaller or larger screen in the facility to fit the camera FoV. In single-lens design, this is achieved by choosing the collimator focal length according to the camera FoV and the screen height. The main drawback of this approach is the necessity of changing - thus purchasing - a new collimator every time a new camera must mounted in the facility. On the contrary, when using a double-lens design, a single screen can be magnified simply by adjusting the relative distance between the relay lens systems and the collimating lens system, leading to the possibility of using the same lenses for multiple cameras without any need to change the hardware. This comes at the cost of distorting and aberrating the light stimuli arriving at the camera pupil when passing through multiple lenses. The two lens systems designed for RETINA are designed to enable variable magnification (thus variable camera FoV), but the lens systems are optimized to minimize the light stimuli warping due to distortions and aberrations. Note that both lens systems are mounted on opto-mechanical stages that enable the alignment and centering of the lens systems with millimeter precision. 

\section{Facility Design and Constraints}\label{sec:design}
In this section, the facility design is outlined by underlining the main drivers and constraints. First, a paraxial lens model is employed to understand the main driver for the facility design. Second, the envelope of the possible facility parameter is established to perform an informed selection of the facility components. Finally, a detailed optical design is performed to optimize the identified preliminary configuration.  

\subsection{The two-lens paraxial model}\label{sec:twolensmodel}
\begin{figure}[ht]
	\centering
	\includegraphics[width=0.8\textwidth]{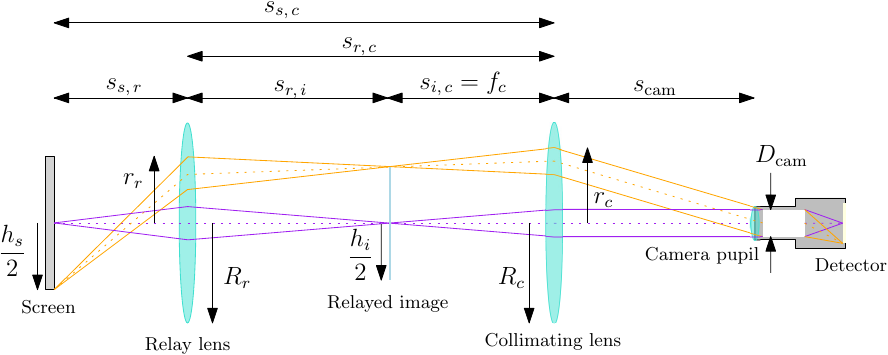}
	\caption{The two-lens paraxial model used for RETINA design.}
	\label{fig:facilityEquation}
\end{figure}

A simplified model for the facility is analyzed in this section to underline the main design drivers and the constraints imposed by the hardware components. To ease the calculations, the paraxial approximation is made. This is a classical assumption in geometrical optics and it has been already used in the past to design optical facilities \cite{panicucci2022tinyv3rse, beierle2019variable, rufino2002laboratory}. Figure~\ref{fig:facilityEquation} shows the two-lens paraxial model that is employed in this section.
\newline
The screen light passes through the relay lens generating the relayed image. The size of the relayed image can be computed as:
\begin{equation}\label{eq:magnificationheight}
	h_i = - M \,h_s
\end{equation}
where $h_i$ is the realy image height, $h_s$ is the screen height, and $M$ is the relay lens magnification. Note that the magnification $M$ correlates with the distance between the screen and its image:
\begin{equation}\label{eq:magnificationdistance}
	s_{r,\,i} = -M s_{s,\,r}
\end{equation}
where $s_{s,\,r}$ is the distance between the screen and the relay lens and $s_{r,\,i}$ is the distance between the relay lens and the relayed image. Recall that the minus sign is justified by the fact that $M<0$ as the generated image is a real one (i.e., the image is not placed on the same size as the real screen with respect to the lens). To achieve this a convex lens must be used as concave lenses have always a positive magnification factor as they generate virtual images (i.e., the image is placed on the same size as the real screen with respect to the lens).\newline
As the light at the camera pupil must be collimated, the distance between the relayed image and the collimating lens $s_{i,\,c}$ must be:
\begin{equation}
	s_{i,\,c} = f_c
\end{equation} 
where $f_c$ is the collimating lens focal length. This ensures that all the rays crossing the collimating lens are parallel.\newline
The optimal working condition for the facility happens when the camera FoV observes the whole screen. As the relayed image is the magnification of the screen image, it is possible to match the relayed image size with the camera FoV as in single lens optical facilities \cite{panicucci2022tinyv3rse,  rufino2002laboratory}:
\begin{equation}\label{eq:FOVmatching}
	h_i = 2f_c\tan\left(\frac{\rm FoV_{cam}}{2}\right)
\end{equation}
where $\rm FoV_{cam}$ is the camera FoV. By using Equations~\ref{eq:magnificationheight} and \ref{eq:FOVmatching}, it is possible to compute the magnification factor necessary to fit the camera in the facility:
\begin{equation}\label{eq:facilityEquation}
	M = -2 \frac{f_c}{h_s}\tan\left(\frac{\rm FoV_{cam}}{2}\right)
\end{equation}
Equation~\ref{eq:facilityEquation} shows that, given some hardware components in the facility (i.e., $f_c$ and $h_s$), it is possible to accommodate different cameras (i.e., $\rm FoV_{cam}$) by changing the magnification level $M$.\newline
Note that, by combining Equations~\ref{eq:facilityEquation} and \ref{eq:magnificationdistance}, it is possible to find a relationship linking the component's location and the camera FoV:
\begin{equation}\label{eq:linkdistances}
	\frac{s_{r,\,i}}{s_{s,\,r}} = 2 \frac{f_c}{h_s}\tan\left(\frac{\rm FoV_{cam}}{2}\right)
\end{equation}
It is worth recalling that the screen and its image are linked by:
\begin{equation}\label{eq:basiclengthequation}
	\frac{1}{f_r} = \frac{1}{s_{s,\,r}} + \frac{1}{s_{r,\,i}}
\end{equation}
where $f_r$ is the relay lens focal length. Thus, by using Equation~\ref{eq:magnificationdistance}:
\begin{equation}\label{eq:distscreen2relay}
	s_{s,\,r} 
	= \left(\frac{M-1}{M}\right)f_r
\end{equation}
\begin{equation}\label{eq:distrelay2image}
	s_{r,\,i} = \left(1-M\right)f_r
\end{equation}
Equations~\ref{eq:distscreen2relay} and \ref{eq:distrelay2image} define the distance between the screen, the relay lens, and the relayed image. Indeed, by knowing the desired level of magnification $M$, the distance between the screen and the relay lens. Note that this also fixes the distance between the relay lens and the relayed image. 

\subsection{Feasible range of components parameters}\label{sec:envelopcomponenetsdesign}
\begin{figure}[ht]
	\centering
	\includegraphics[width=0.6\textwidth]{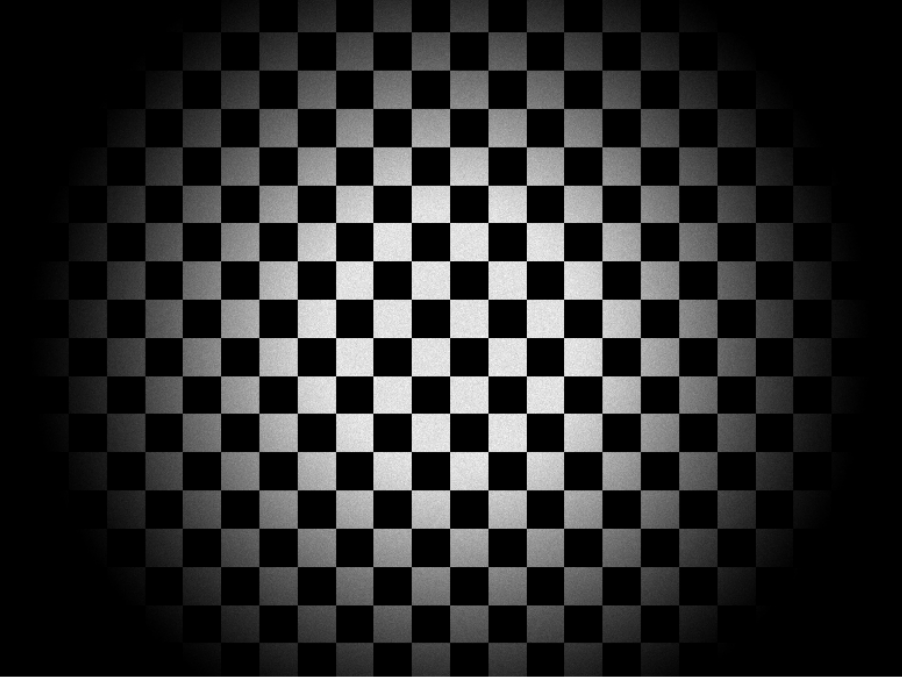}
	\caption{An example of vignetting when the vignetting constraint is not verified.}
	\label{fig:vignetting}
\end{figure}

\begin{figure}[ht]
	\centering
	\begin{subfigure}{0.47\textwidth}
		\includegraphics[width=\textwidth]{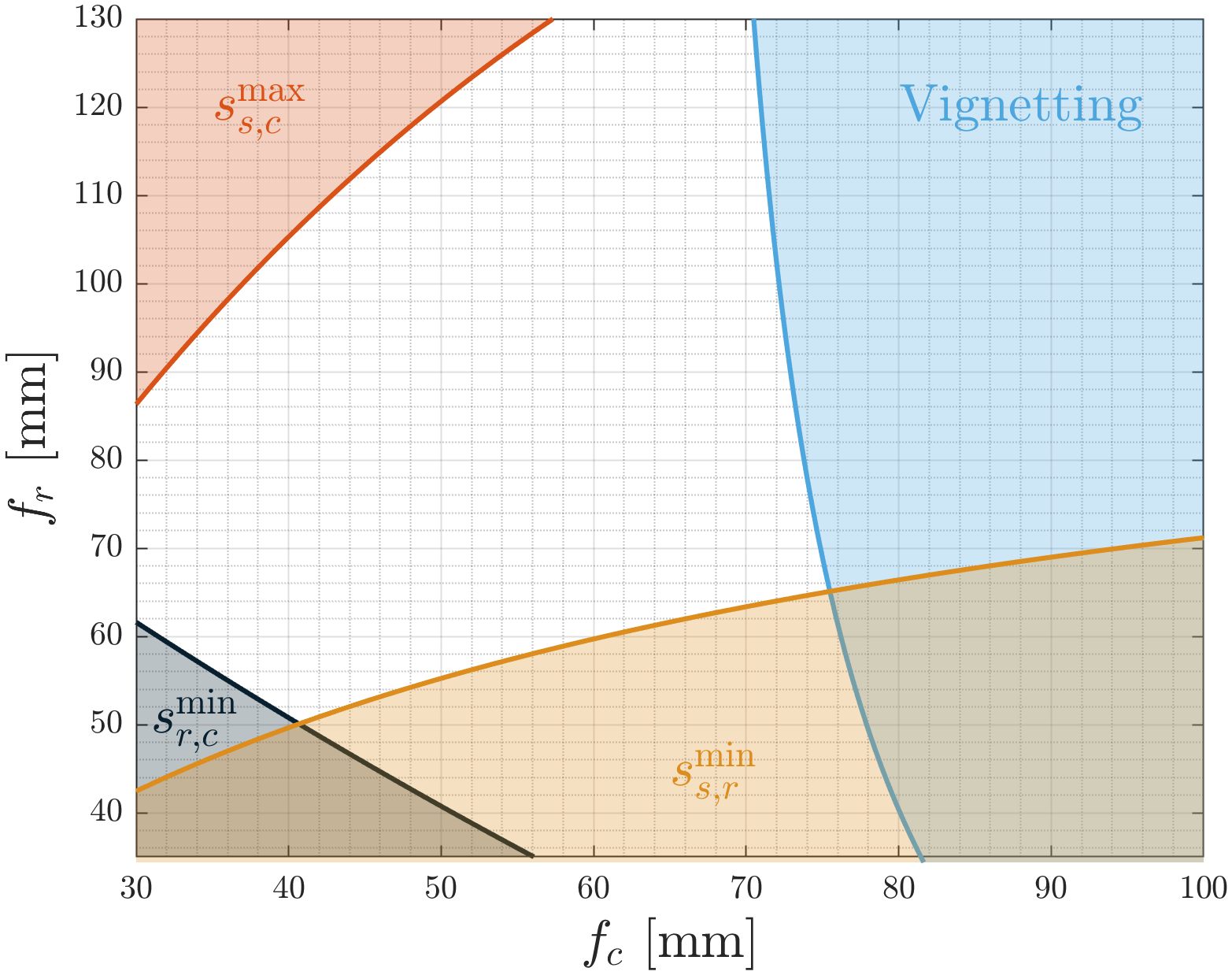}
		\caption{$R_c = 1''$, $R_r = 1''$}
		\label{fig:feasRange2inch}
	\end{subfigure}
	\hfill
	\begin{subfigure}{0.47\textwidth}
		\includegraphics[width=\textwidth]{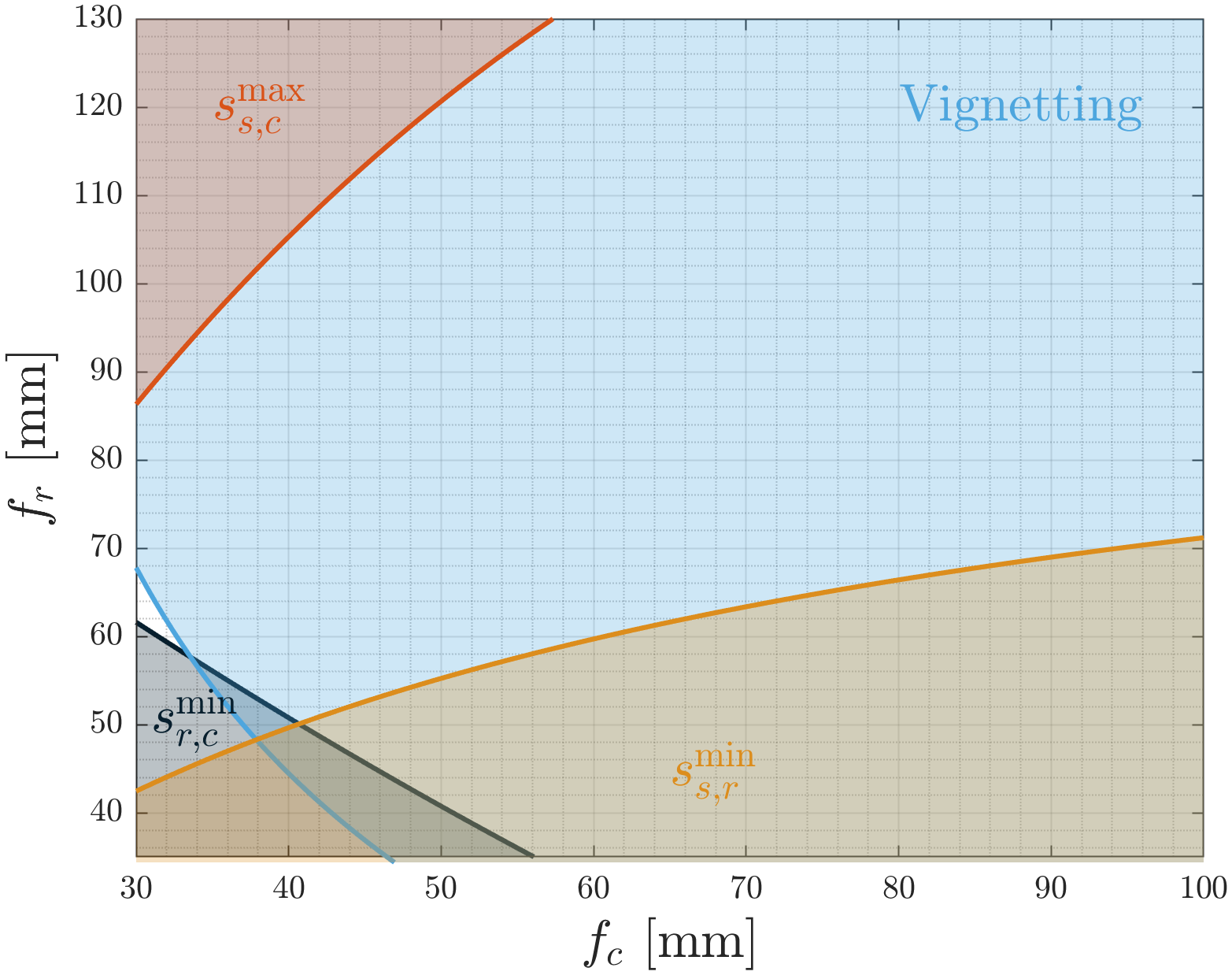}
		\caption{$R_c = 0.5''$, $R_r = 1''$}
		\label{fig:feasRange1inch}
	\end{subfigure}
	\caption{The range of feasible focal lengths for different values of lens radii.}
	\label{fig:feasRange}
\end{figure}

From the paraxial model in Section~\ref{sec:twolensmodel}, it is possible to understand that three variables are driving the preliminary design of the facility: $f_c$, $f_r$, and $h_s$. It is worth noting that only one free parameter exists as the components properties are linked by Equations~\ref{eq:linkdistances} and \ref{eq:basiclengthequation}. For the design of RETINA, the screen was selected among the ones available in the market for its illumination dynamical range, its reduced size, and its high frame rate (i.e., 60 Hz). This implies that the constraints given by Equations~\ref{eq:linkdistances} and \ref{eq:basiclengthequation} define a range of feasible focal lengths to fulfill the facility requirements on the operative camera FoV. This section aims to present this feasible domain and to present the selected preliminary solution.\newline
The first constraint to be taken into account is the total length of the rails where the opto-mechanical stages are mounted. The maximum distance between the components mounted in the facility is:
\begin{equation}
	s_{s,\,c} = s_{s,\,r} + s_{r,\,c} = s_{s,\,r} + s_{r,\,i} + s_{i,\,c} = - f_{r} \left(\frac{1 + M^2}{M}\right) + f_c \leq s_{s,\,c}^{\rm max} 
\end{equation}  
where $s_{s,\,c}^{\rm max} = 850$ mm is the maximum length allowable in RETINA. Note that $M$ is computed by Equation~\ref{eq:facilityEquation}. This constraint is necessary to ensure that all components are correctly accommodated within the enclosing box and the correct movement of the stages on the transnational rail.\newline
The second constraint to consider is the minimal allowed distance between the opto-mechanical stages. Indeed, owing to their physical size, it is not possible to approach two stages more than a given threshold $s_{r,\,c}^{\rm min} = s_{s,\,r}^{\rm min} = 100$ mm. Therefore:
\begin{equation}
	\left(1-M\right)f_r + f_c \leq s_{r,\,c}^{\rm min}
\end{equation}
\begin{equation}
	\left(\frac{M-1}{M}\right)f_r \leq s_{s,\,r}^{\rm min}
\end{equation} 
The last constraint relates to vignetting. Indeed, when observing through the lenses, all the rays entering the camera pupils must come from the screen. When some rays are not passing through both lenses, vignetting is present. When this happens, the camera does not receive all the rays emitted by the screen, leading to a loss of stimulation at the image border. An example of this effect is reported in Figure~\ref{fig:vignetting}. 
\newline
To evaluate the effect of vignetting in the stimulation, a ray casting procedure is developed. Let $R_r$ and $R_c$ be the radius of the relay and the collimating lenses respectively. Let $d_{\rm cam}$ be the camera pupil as depicted in Figure~\ref{fig:facilityEquation}. In the casting procedure, the rays having the same inclination of the camera FoV are considered as these are the outermost rays entering the camera pupil. These rays are the ones depicted in orange in Figure~\ref{fig:facilityEquation}. These rays are cast through the paraxial model to compute their vertical intersection with the two lenses. The intersection with the collimating lens is labeled $r_c$, whereas the one with the relay lens is $r_r$. Note that, given a camera FoV, it is crucial to cast rays from the camera pupil limit to correctly assess vignetting as the intersection of the upper or lower ray could not be the extremal intersection with the lenses. No vignetting occurs when $r_r \leq R_r$ and $r_c \leq R_c$. When both conditions occur, the extremal rays of the camera FoV are correctly cast from the camera pupil to the screen. In the vignetting analyses, the camera pupil $d_{\rm cam}$ is set to 10 mm (i.e., F\# = 1.2 for a 12 mm focal length optical head), and the distance between the camera pupil and the collimating lens (i.e., $s_{\rm cam}$) is the optimal one for the considered camera FoV, given a combination of lenses radius (i.e., $R_r$ and $R_c$) and focal lenses (i.e., $f_r$ and $f_c$). To do so, before performing the vignetting analyses, a look-up table is computed for $s_{\rm cam}$ to be used in the vignetting analysis to assess the feasibility of a given FoV.\newline
According to these constraints, a series of analyses is performed by varying $f_c$ and $f_r$ to understand whether the facility can accommodate a camera FoV fulfilling the requirements. As the vignetting analysis is dependent on the lens radii as well, two different simulations were performed by considering that the collimating lens radius could assume the value of 1" or 0.5", which is compliant with COTS lens radii available on the market. Moreover, it is worth mentioning that cemented triplet lenses provide high performance against aberrations with reduced complexity in optical design. As these lenses are only available as COTS with a radius of 0.5", it is of interest to investigate a smaller relay lens radius. Note that the relay lens radius is fixed at 1" because a smaller lens radius would have been too stringent in terms of vignetting, providing no feasible solution.\newline
The results of these analyses are shown in Figures~\ref{fig:feasRange2inch} and \ref{fig:feasRange1inch} for $R_c$ equal to 1" and 0.5" respectively. The black and brown areas are associated with unfeasible solutions due to the interdistances between the optical components. Indeed, too small focal lengths would imply small distances between components. The red area is associated with a total length greater than the maximum. This is associated with low relay lens focal lengths and small FoVs. Lastly, the blue area is associated with the vignetting effect. This effect limits the collimator focal lengths as the ideal image must be further away from the lens, leading to some rays not falling into the relay lens. The range of feasible solutions for the design is the not colored one which provides all the couples $\left(f_r, \, f_c\right)$ that are geometrically compliant with the constraints.
\newline
Moreover, Figure~\ref{fig:feasRange1inch} shows a smaller range of feasible focal lengths, implying that a larger collimating lens radius gives more variability in the lens selection. Moreover, it is worth noting that for a greater camera pupil diameter, a small collimating lens diameter could imply the constant presence of vignetting. Therefore, RETINA was designed with $R_c = R_r = 1''$. Moreover, the selected focal lengths must be located in the upper region of the feasible areas as higher focal lengths are preferred to achieve lower magnification errors when the components are not correctly placed in the facility \cite{beierle2019variable}.

\subsection{Low-aberration optical design}\label{sec:lowaberrationsdesign}
\begin{figure}[t]
	\centering
	\centering
	\begin{subfigure}{0.95\textwidth}
		\centering
		\includegraphics[trim={0cm 5cm 0cm 41cm}, clip, width=\columnwidth]{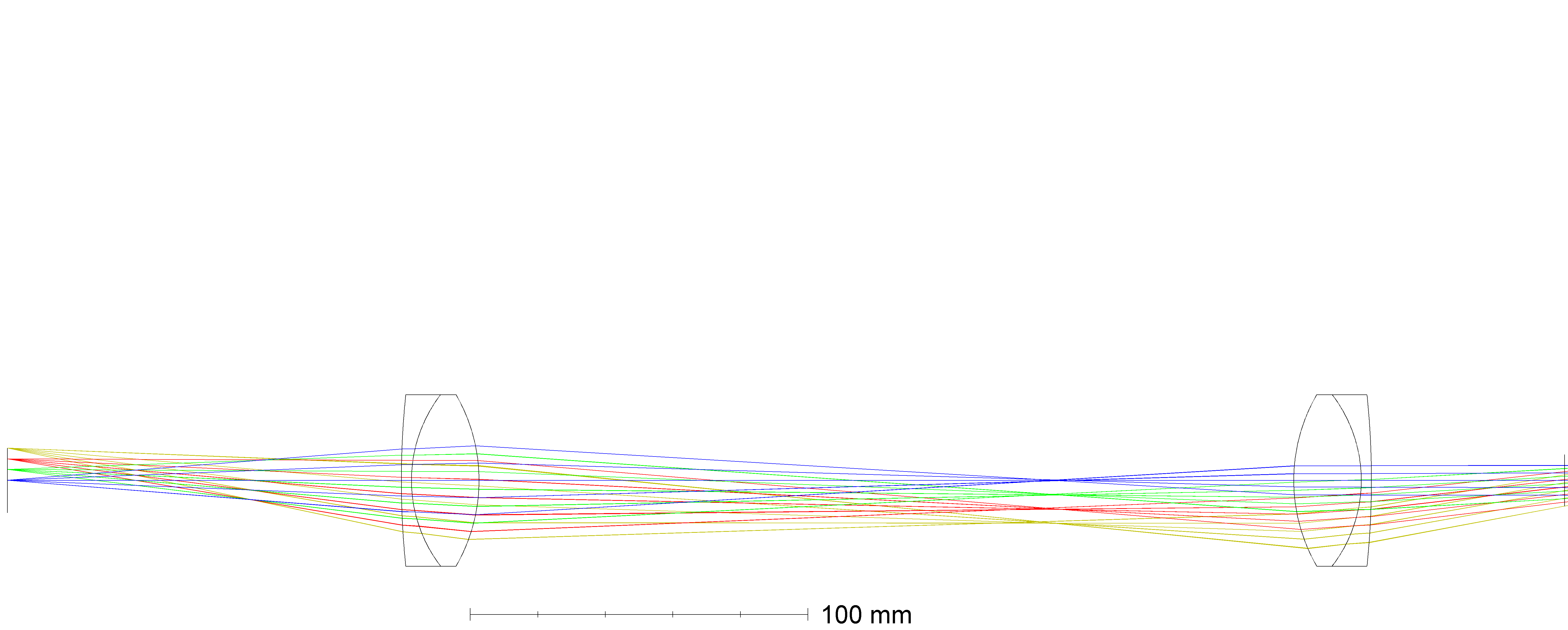}
		\caption{25-mm objective.}
		\label{fig:initialConfig_25_layout}
	\end{subfigure}\\
	\begin{subfigure}{0.95\textwidth}
		\centering
		\includegraphics[trim={0cm 4cm 0cm 41cm}, clip, width=\columnwidth]{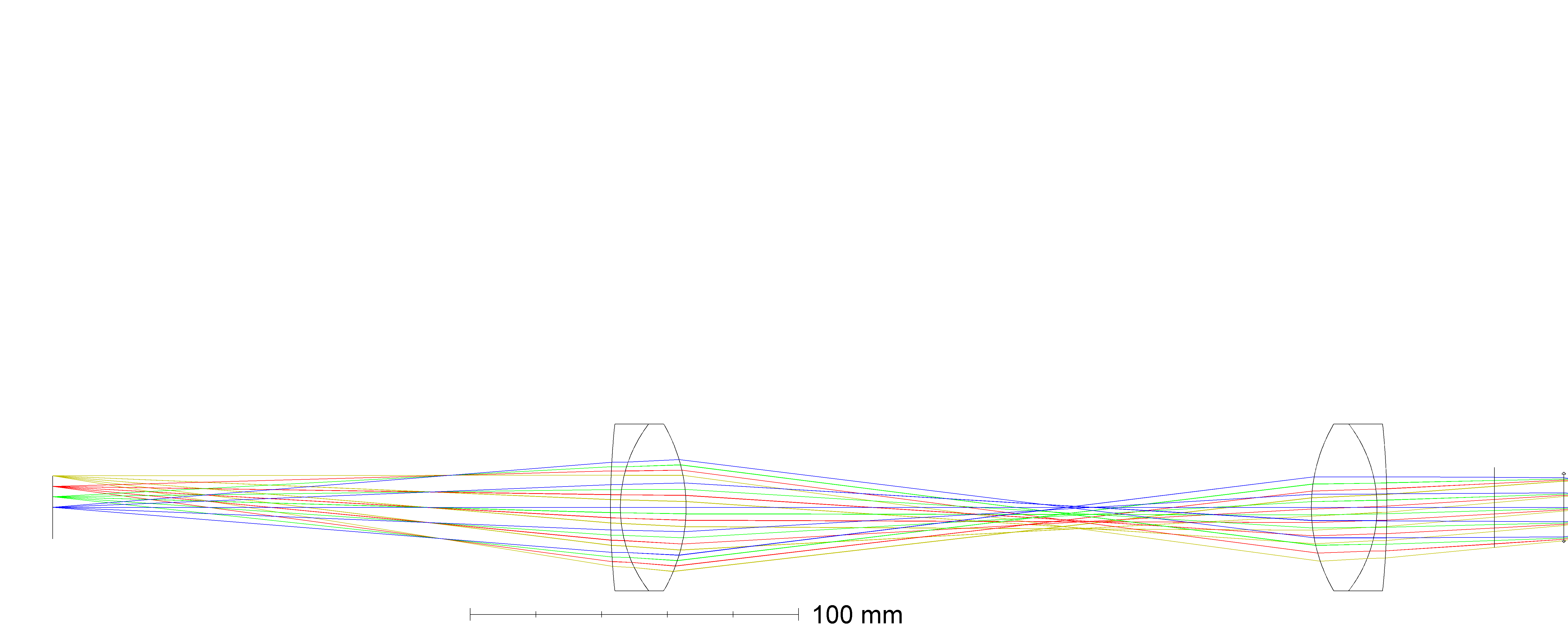}
		\caption{50-mm objective.}
		\label{fig:initialConfig_50_layout}
	\end{subfigure}
	\caption{Zemax sketch of the first investigated solution. Ray colors are associated with the screen-emitting height.}
	\label{fig:firstIteration}
\end{figure}

\begin{figure}[t]
	\centering
	\begin{subfigure}{0.95\textwidth}
		\centering
		\includegraphics[trim={15cm 5cm 0cm 41cm}, clip, width=\columnwidth]{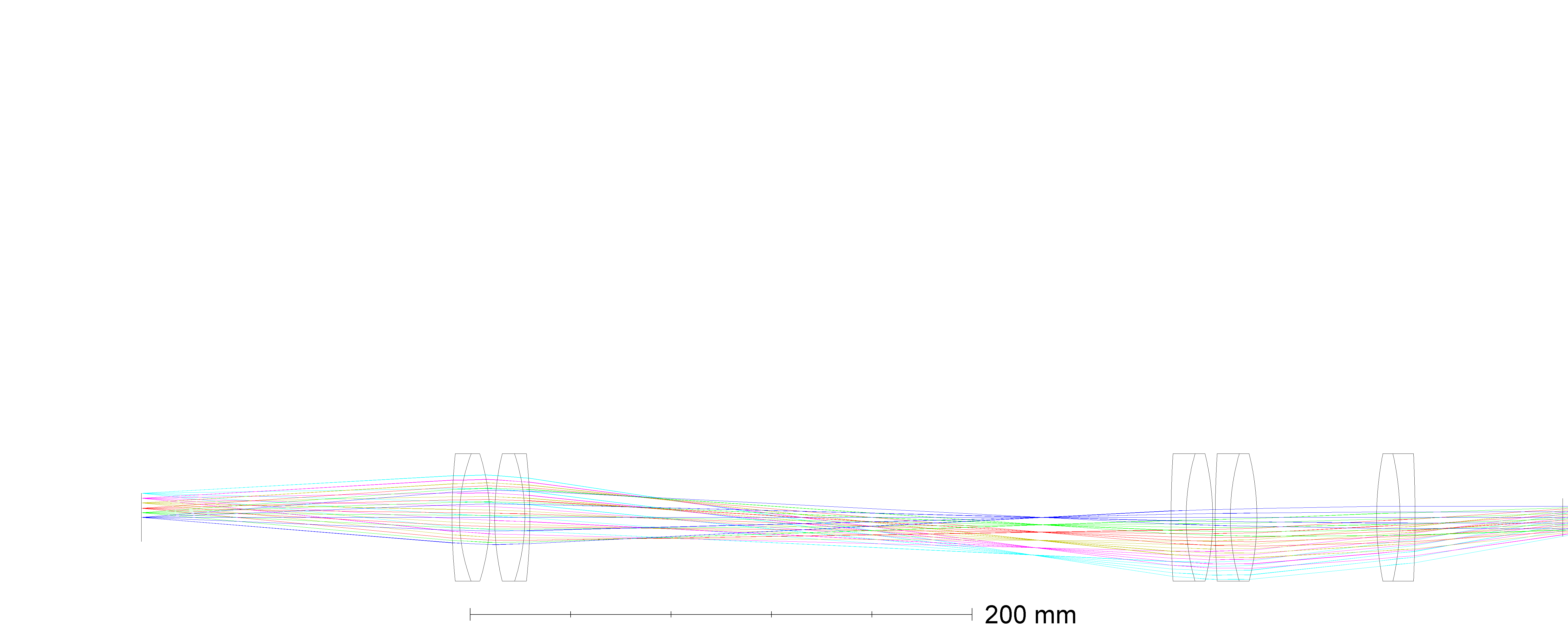}
		\caption{25-mm objective.}
		\label{fig:RETINA_25_layout}
	\end{subfigure}\\
	\begin{subfigure}{0.95\textwidth}
		\centering
		\includegraphics[trim={15cm 4cm 0cm 41cm}, clip, width=\columnwidth]{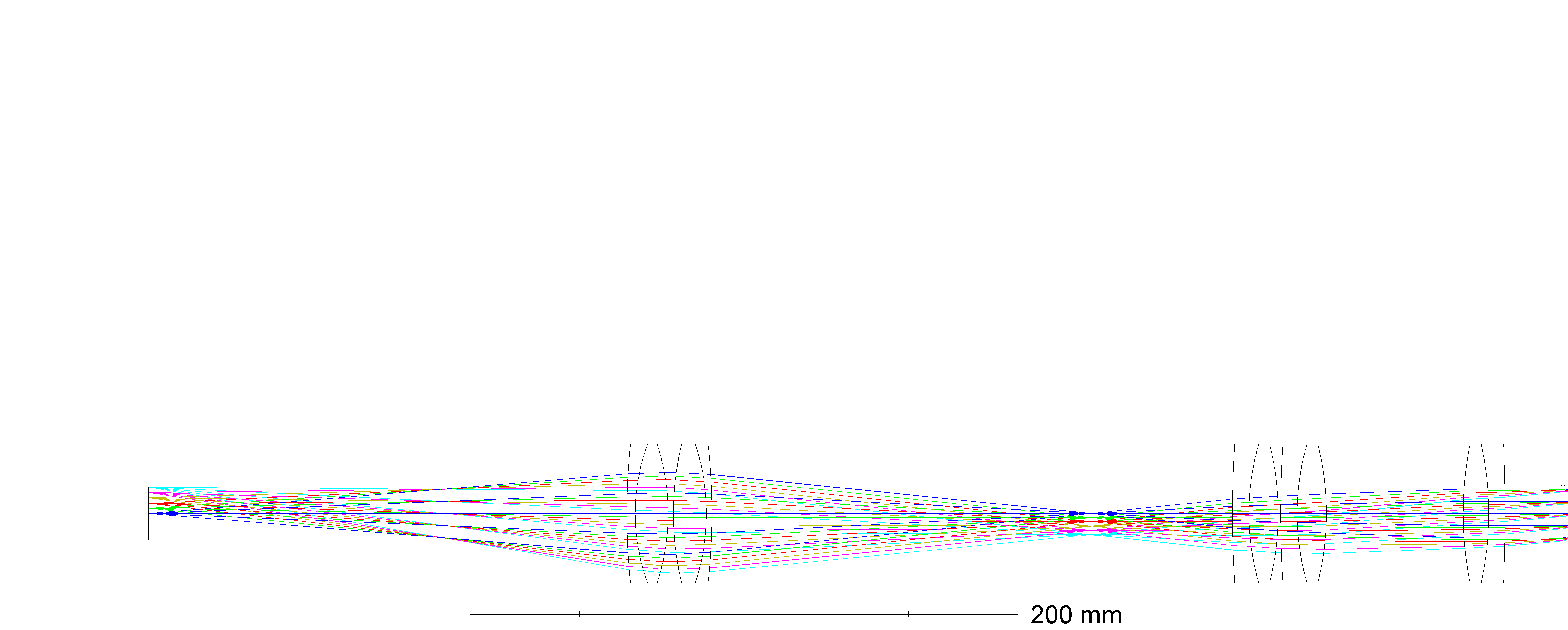}
		\caption{50-mm objective.}
		\label{fig:RETINA_50_layout}
	\end{subfigure}
	\caption{Zemax sketch of the final design. Ray colors are associated with the screen-emitting height.}
	\label{fig:finalDesign}
\end{figure}

\begin{figure}
	\centering
	\begin{subfigure}[h]{0.8\textwidth}
		\centering
		\includegraphics[width=\textwidth]{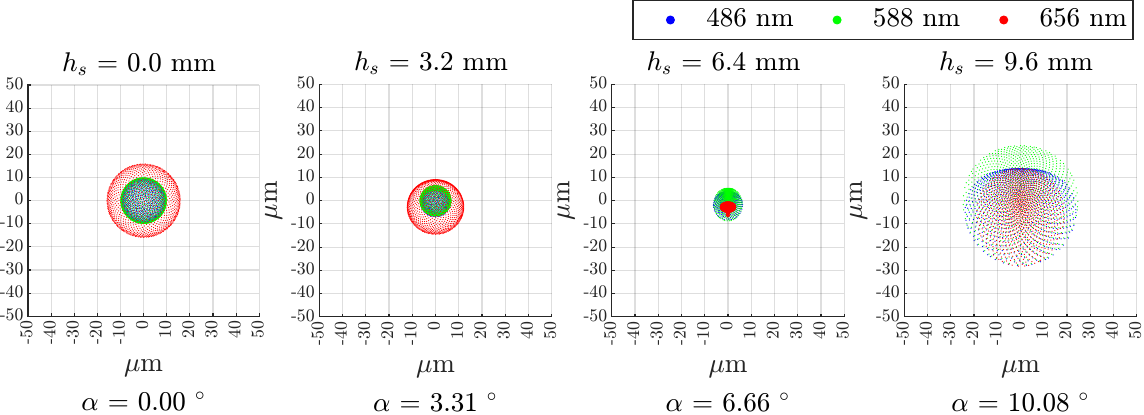}
		\caption{25 mm}
		\label{fig:finalDesignSpotDiagram25mm}
	\end{subfigure}\\
	\begin{subfigure}[h]{0.8\textwidth}
		\centering
		\includegraphics[width=\textwidth]{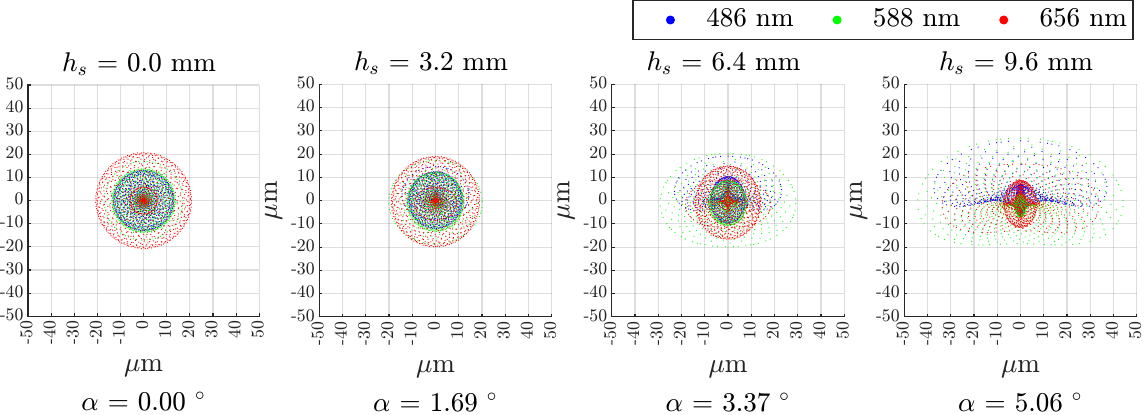}
		\caption{50 mm}
		\label{fig:finalDesignSpotDiagram50mm}
	\end{subfigure}
	\caption{Spot diagram of the model reported in Figures~\ref{fig:RETINA_25_layout} and \ref{fig:RETINA_50_layout} at F/2.8 aperture.}
	\label{fig:finalDesignSpotDiagram}
\end{figure}

\begin{figure}[t]
	\centering
	\includegraphics[width=0.8\textwidth]{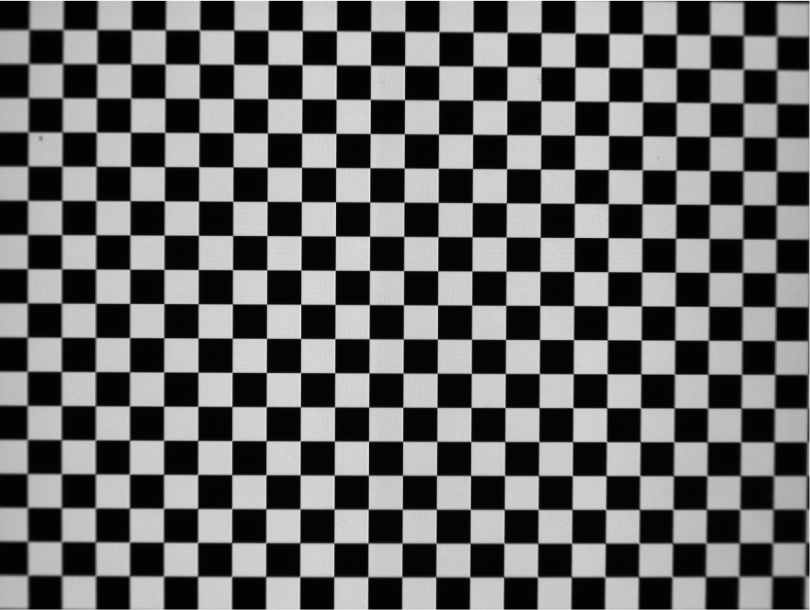}
	\caption{An image of a regular pattern acquired in RETINA with the low-aberration configuration.}
	\label{fig:finalDesignImageBlur}
 \end{figure}

\begin{figure}[ht]
	\centering
	\includegraphics[width=0.7\textwidth, trim={0cm 0.1cm 0cm 0cm}, clip]{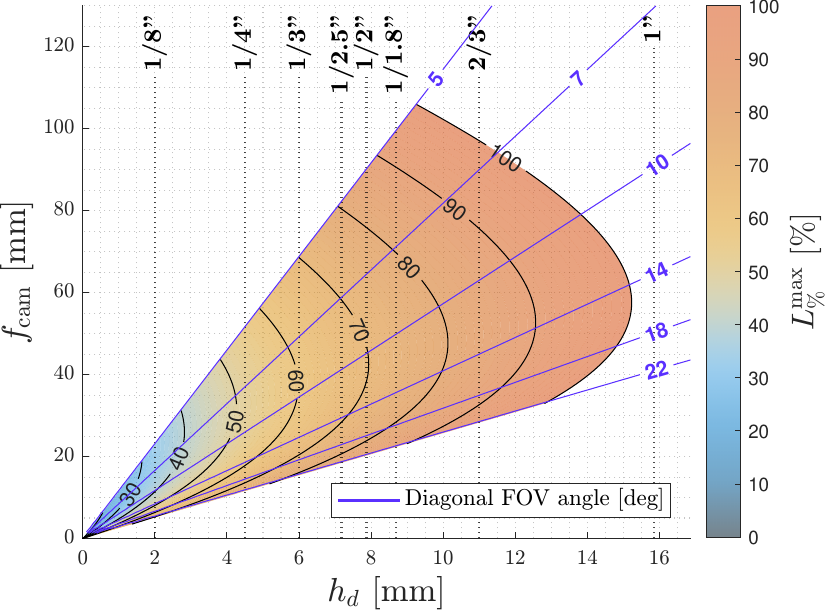}
	\caption{The envelope of cameras that can be accommodated in the facility along with the optical performance index for each camera.}
	\label{fig:feasibleCameras}
\end{figure}

In this section the optical performance of the facility is studied and the lens systems are optimized to remove aberrations from the preliminary design. To do so, a dedicated study in Zemax OpticStudio is performed as a simple paraxial model cannot predict the optical distortions and aberrations induced by the lenses mounted in the facility. The software is the state-of-the-art for optical design and analysis and it can compute the optical performance of an optical system by sequential raytracing. As the optical design software requires the setting of the entrance camera pupil to compute the optical performance, the camera considered in this section has a F- number of 2.8.  All the lenses considered in this sections are COTS components to avoid cost increase and complex lens design. The lenses are modeled in Zemax according to their data-sheet characteristics to correctly assess RETINA optical performance. Note that the use of COTS components make the optical performance optimization a challenging task as most of the optical systems optimize the lenses characteristics (e.g., lens shape and curvature) to reduce optical aberrations. In the case of COTS components this is not possible as the only design variable is the type of lens to be purchased and their relative distance. Therefore, this may result in several iterations to determine which is the correct configuration to be selected. This section will not detail all the iterative process that led to the final solution, but it presents one of the first investigated solutions to show the importance of optical performance assessment. Then, the final configuration is presented.\newline
The first investigated solution considers two 1"-radius achromatic cemented doublets with focal lengths of 75 mm. These lenses are selected as they reduce chromatic aberrations with a cemented double glass layer. In a preliminary investigation, also Hastings triplets were investigated owing their superior optical performance, but they were discarded as their radius is 0.5" for COTS components (see Figures \ref{fig:feasRange2inch} and \ref{fig:feasRange1inch}). The Zemax model of this first iteration is reported in Figures~\ref{fig:initialConfig_25_layout} and \ref{fig:initialConfig_50_layout} for an objective of 25 mm and 50 mm respectively.
\begin{figure}[t]
	\centering
	\begin{subfigure}[h]{0.8\textwidth}
		\centering
		\includegraphics[width=\textwidth]{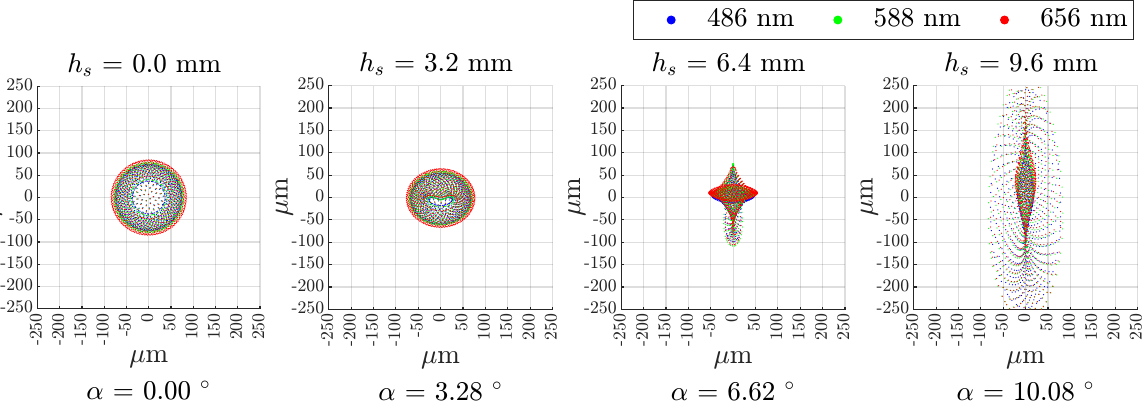}
		\caption{25 mm}
		\label{fig:spot25_firstDesign}
	\end{subfigure}\\
	\begin{subfigure}[h]{0.8\textwidth}
		\centering
		\includegraphics[width=\textwidth]{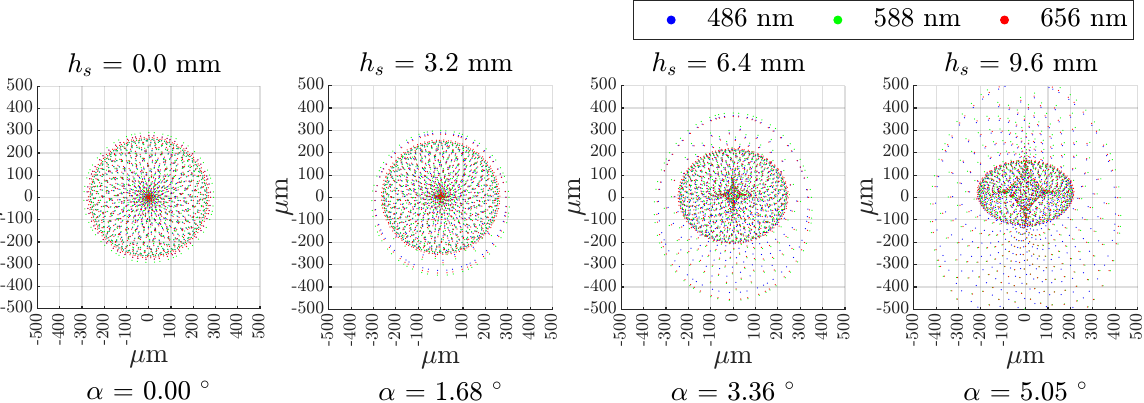}
		\caption{50 mm}
		\label{fig:spot50_firstDesign}
	\end{subfigure}
	\caption{Spot diagram of the model reported in Figures \ref{fig:initialConfig_25_layout} and \ref{fig:initialConfig_50_layout}.}
	\label{fig:firstIterationSpotDiagram}
\end{figure}
\newline
On the left part of the model, the screen generates ray beams at different distances from the optical center to assess the optical response of the system. The collimator model is downloaded from the lens producer website and imported into the Zemax model. On the right part of the model, a paraxial objective was created to emulate the optical properties of a camera. This was done to evaluate the distortions and aberrations of the RETINA lenses apparatus alone. The camera is devised with two lenses and an aperture stop to correctly fit the camera focal length and the camera pupil aperture. In Figures~\ref{fig:initialConfig_25_layout} and \ref{fig:initialConfig_50_layout}, the ray beams are reported with different colors according to their distance with respect to the screen center for 25-mm and 50-mm objectives respectively. Each ray beam is propagated through the system to impact the camera sensor plane. This defines the spot diagram shown in Figures~\ref{fig:spot25_firstDesign} and \ref{fig:spot50_firstDesign} where the spot color is associated with the ray wavelength (i.e., blue=486 nm, green=588 nm, and red=656 nm) for 25-mm and 50-mm objectives. 
\begin{figure}[t]
	\centering
	\includegraphics[width=0.8\textwidth]{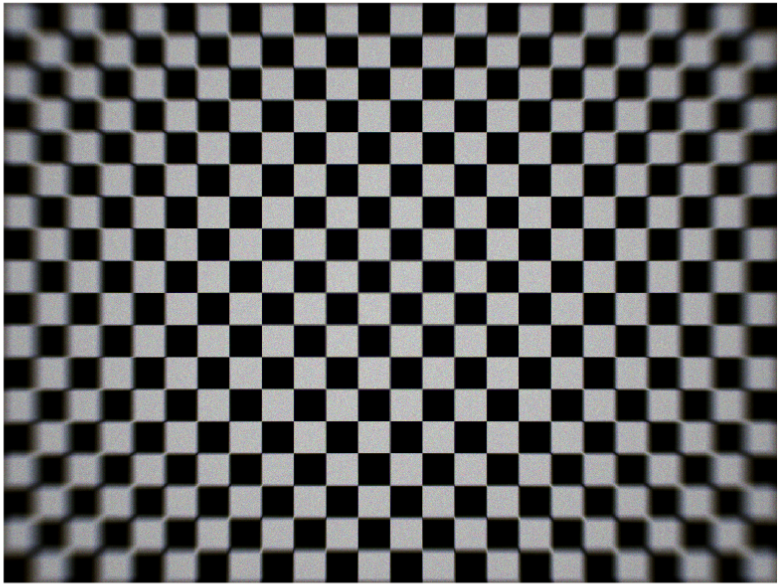}
	\caption{An image of a regular pattern acquired in RETINA with the first investigated configuration.}
	\label{fig:firstIterationImageBlur}
\end{figure}
In Figures~\ref{fig:spot25_firstDesign} and \ref{fig:spot50_firstDesign} $h_s$ is the point on the detector generating the spot diagram and $\alpha$ is the angular location associated with the center of the spot diagram. Note that the ray beam locations on the screen are selected to impact at different location of the detector to account for variation of the performance in off-axis conditions. The numerical assessment of the simulation is reported in Table~\ref{tab:firstIterationSpotDiagramTable} where the RMS radius (i.e., the root-mean-square value of the radius of all the rays generated from a single ray beam) is provided in micrometer as a function of the distance from the detector center. The micrometer are chosen as camera pixels are usually of these order of magnitude, leading to an easy comparison according to the selected detector.
\begin{table}[ht]
	\centering
	\caption{RMS radius of the spot diagram as a function of the detector position for the model reported in Figure~\ref{fig:firstIteration}}
	\label{tab:firstIterationSpotDiagramTable}
	\begin{tabular}{c|cccc}
		Camera & Position on & Position on & Position in & RMS \\
		Setup &  the screen [mm] &   the detector [mm] &   the FoV [deg]& radius [$\mu$m]\\
		\hline
		\multirow{4}{*}{25-mm objective}
		& 0 & 0 & 0  & 62.73\\
		& 3.2 & 1.43 & 3.27  & 50.63\\
		& 6.4 & 2.89 & 6.61 & 47.43\\		
		& 9.6 & 4.41 & 10.08 & 128.34\\
		\hline
		\multirow{4}{*}{50-mm objective}
		& 0  & 0  & 0  & 219.27\\
		& 3.2  & 1.47  & 1.68  & 224.31\\
		& 6.4  & 2.94  & 3.36  & 246.02\\		
		& 9.6  & 4.41  & 5.05 & 298.93\\
	\end{tabular}
\end{table}
It is possible to conclude that rays passing through the lenses close to the optical axis experience low aberrations and distortions, while the external ones show a very large degree of aberration. In particular, ray are extremely spread, leading to a severe blurring effect at the corner of the detector when observing the screen. An example of an image taken with this configuration is reported in Figure~\ref{fig:firstIterationImageBlur}, proving experimentally the studied setup.\newline
In order to mitigate the aberrations and effects related with blurring, a different optical design must be put in place. As COTS components are used, it is not possible to optimize the lens shape and surface properties to minimize the aberrations induced by the collimating and relay lenses. A possible solution is to use a higher number of lenses, leading to an increase of degree of freedom to increase the optical performance \cite{madrid2023off,ryzhikov2022method}. Indeed, a higher number of lenses implies a higher number of distance to be optimized. It is worth recalling that a lens system composed of two lenses has an equivalent focal length $f_{\rm equiv}$ given by:
\begin{equation}\label{eq:eqFocalLength}
\frac{1}{f_{\rm equiv}} = \frac{1}{f_1} +  \frac{1}{f_2} -  \frac{d}{f_1\,f_2}
\end{equation}
where $f_1$ and $f_2$ is the focal length of the first and second lens composing the lens system, and $d$ is the distance between the two lenses. In the case of multiple lenses, Equation~\ref{eq:eqFocalLength} can be used iteratively to compute the equivalent focal length. For both the collimating and relay lens system the equivalent focal lengths must be compliant with the feasible ranges reported in Section~\ref{sec:envelopcomponenetsdesign}, leading to design constraints when selecting the focal length of each single lens.\newline
The selection of the number of lenses, their focal length, and their relative distance to ensure optical performance while ensuring the previous constraints has been performed by simulations and iterations in Zemax. The main driver used in the selection process are:
\begin{enumerate}
	\item The relay and collimating lens systems do not induce a high field curvature for the considered wavelengths (i.e., red, blue and green) on the tangential and sagitarial planes. This condition ensures that the screen image as seen from the camera is not warped too much during the propagation through the lens systems. 
	\item The equivalent single-lens focal lengths for both lens systems must be compliant with the feasible region in Figure~\ref{fig:feasRange2inch}, with a preference for higher focal lengths.
\end{enumerate}
Note that the above conditions were studied for narrow and wide FoV so to ensure acceptable optical performance for the two extremal cases of the facility design. The final solution is shown Figures~\ref{fig:RETINA_25_layout} and \ref{fig:RETINA_25_layout} for wide and narrow FoVs and it is composed of 5 lenses. In particular, the collimating lens is composed of two lenses generating an equivalent focal length of 86.8 mm and the relay lens is composed of three lenses providing an equivalent focal length of 82.8 mm. The optical performance of the final design are assessed by studying RETINA spot diagram for different objective focal length at f/2.8. Note that this value of the aperture is selected as it is consistent with star tracker typical values and as larger apertures are usually associated with larger aberrations and distortions. The spot diagrams are reported for three wavelengths in Figures~\ref{fig:finalDesignSpotDiagram25mm} and \ref{fig:finalDesignSpotDiagram50mm} for 25-mm and 50-mm objectives respectively.
\begin{table}
	\centering
	\caption{RMS radius of the spot diagram as a function of the detector position for the model reported in Figure~\ref{fig:finalDesign}}
	\label{tab:finalDesignSpotDiagramTable}
	\begin{tabular}{c|cccc}
		Camera & Position on & Position on & Position in & RMS \\
		Setup &  the screen [mm] &   the detector [mm] &   the FoV [deg]& radius [$\mu$m]\\
		\hline
		\multirow{4}{*}{25-mm objective}
		& 0 & 0 & 0  & 10.38\\
		& 3.2 & 1.45 & 3.31  & 7.71\\
		& 6.4 & 2.91 & 6.6 & 5.04\\		
		& 9.6 & 4.41 & 10.07 & 17.34\\
		\hline
		\multirow{4}{*}{50-mm objective}
		& 0  & 0  & 0  & 12.04\\
		& 3.2  & 1.47  & 1.69  & 11.82\\
		& 6.4  & 2.94  & 3.37  & 12.61\\		
		& 9.6  & 4.41  & 5.05 & 17.59\\
	\end{tabular}
\end{table}
\newline
The performance is improved when compared with the simulations reported in the first investigated solution thanks to the low aberration design. Clearly, the performance degrades by increasing the angular distance with respect to the boresight for both considered objectives, but the influence of distortions and aberrations remain limited. In particular, Table \ref{tab:finalDesignSpotDiagramTable} that the spot diagram is below the usual size of a single pixel, implying that light coming from a single point on the screen would be collected by a single pixel in the camera. An example of an image taken with the low-aberration set up is reported in Figure~\ref{fig:finalDesignImageBlur}. It is possible to note optical performance is improved, in particular at the image edges, implying precise stimulation of the camera along all its filed of view.
As RETINA has been designed taking only into account the camera FoV, a study has been performed to understand which camera focal lengths $f_{\rm cam}$ and the camera diagonal sizes $h_d$ can be used in the facility. This defines the envelope of camera characteristics that can be used in RETINA. Moreover, to assess the quality of the stimulation for each camera, the following optical performance index parameter is introduced:
\begin{equation}
	L_{\%}^{\max} = 100\, \max\left(\frac{r_r}{R_r},\,\frac{r_c}{R_c}\right)
\end{equation} 
The optical performance index gives information about the quality of the stimulation when a camera is mounted in RETINA. To quantify it, the highest crossing of the extremal ray cast in the vignetting analysis is used. Indeed, the optical performance of a lens system degrades when rays pass at the edges of a lens, as the paraxial hypothesis is less valid in this area leading to higher aberrations and distortions. Therefore, the higher $L_{\%}^{\max}$ is, the more the optical performance is degraded as light rays pass at the edge of the lens systems. It is worth noting that optical degradation does not imply that the image cannot be used for navigation. The degradation is evaluated with respect to the nominal performance that is computed in the best conditions possible (i.e., an optical ray passing at the center of all lenses where the paraxial hypothesis holds). Figure~\ref{fig:feasibleCameras} shows the envelope of the camera characteristics parameters along with the optical performance index considering a camera with F\#=2.8. Vertical dashed lines are associated with standard sensor diagonal size for space and Earth detectors, while purples lines denote the camera FoVs. 

\section{Optical Facility Geometrical Calibration}\label{sec:calibrationalgorithm}
To correctly stimulate the camera, the facility must be able to correctly project and reproduce the observed scene as if the camera would be place on orbit. In theory, this can be achieved by limiting the lens distortions and aberrations, by centering the facility components, and by correctly aligning all the components. In practice, lenses induce distortions and aberrations errors that must be cleaned out as not due to camera objective. Moreover, facility components are not perfectly aligned and centered leading to the error in the observed scene that prevent to validate and characterize VBN and IP algorithms. This implies that a calibration procedure must be put in place in order to characterize the the facility geometrical errors and to compensate them at software level~\cite{panicucci2021autonomous}.

\subsection{Facility Projection Model}
\begin{figure}[t]
	\centering
	\includegraphics[width=0.8\textwidth]{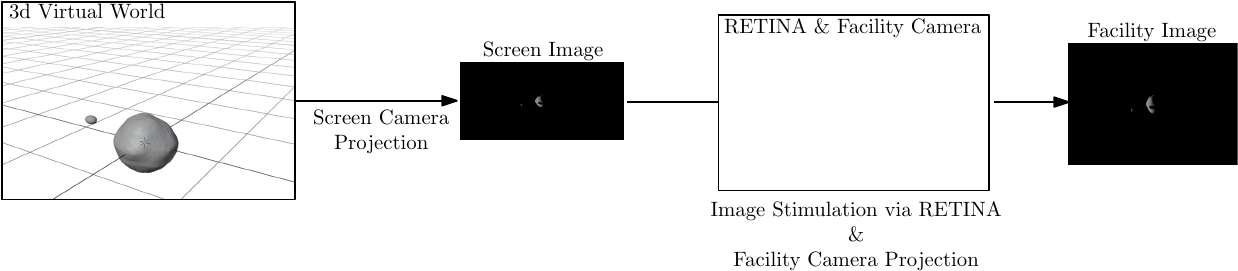}
	\caption{The RETINA workflow from the 3D virtual environment to the image acquired in the facility.}
	\label{fig:RETINAworkflow}
\end{figure}

\begin{figure}[t]
	\centering
	\includegraphics[width=\textwidth]{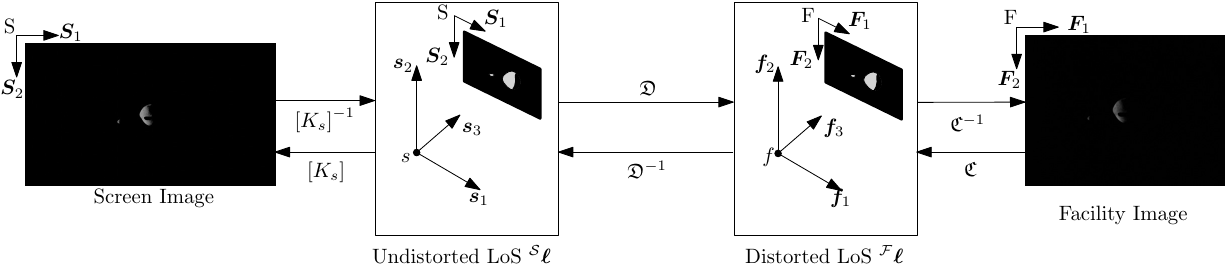}
	\caption{The RETINA workflow from the 3D virtual environment to the image acquired in the facility.}
	\label{fig:projectionAndDistortionRelationship}
\end{figure}

\begin{figure}[b]
	\centering
	\includegraphics[width=\textwidth]{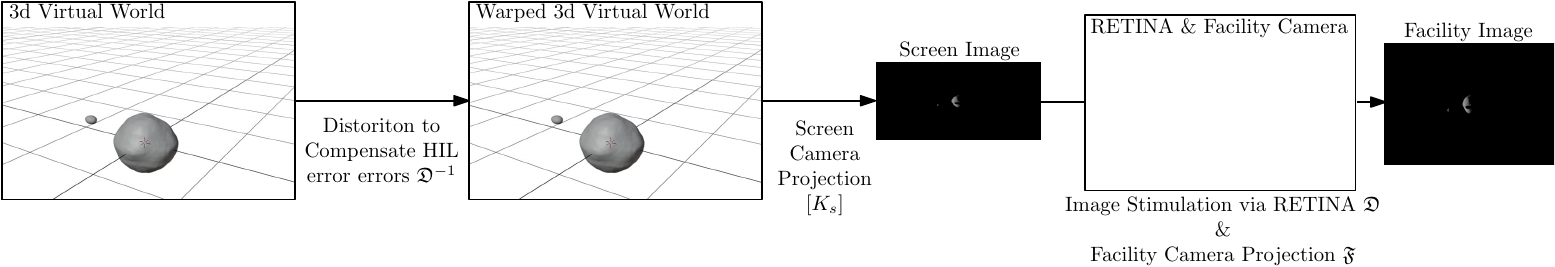}
	\caption{The upstream compensation used to ensure the equivalence of the LoSes in the 3D virtual world with the ones detected by the camera.}
	\label{fig:upstreamCompensation}
\end{figure}

To estimate the facility distortions and misalignment, it is necessary to establish a fruitful projection model to be used for error quantification and compensation. More in general, the goal of this section is to gather a projection model that can ensure that the camera in the facility observes the same LoSes as if it would be placed in the 3D virtual rendering engine \cite{panicucci2022tinyv3rse}. This procedure is detailed in \citet{panicucci2022tinyv3rse} and a summary is provided here for the sake of completeness.\newline
An overview of the RETINA workflow modeled in this section is presented in Figure~\ref{fig:RETINAworkflow}.
In the following, it is assumed in the projection model that the camera mounted in RETINA is already calibrated for what concerns the distortions and intrinsic camera matrix. This implies that it is possible to gather undistorted homogeneous line of sights (LoSes) from distorted pixel coordinates. Therefore, it is necessary to establish a mapping from the 3D virtual world to the screen image and to determine an error model for RETINA. Once these two contributions are modeled, it is possible to estimate the RETINA error model with a dedicated calibration procedure.\newline
Let \sdr{s} and \sdr{f} be the 2D reference frames associated with the screen and facility images, respectively. Let \SdR{s} and  \SdR{f} be the 3D reference frames associated with the screen and facility cameras, respectively. The projection from the 3D virtual world to the screen image can be represented by a linear projection matrix $\left[K_s\right]$ associated with the screen camera (i.e., the camera that generate the images to be displayed on the screen via a rendering tool). The screen camera projection matrix  $\left[K_s\right]$ is expressed as follows:
\begin{equation}
	 \left[K_s\right] = \begin{bmatrix}
	 	f_{s_1} & 0 & c_{s_1} \\ 0 & f_{s_2} & c_{s_2} \\ 0 & 0 & 1
	 \end{bmatrix}
\end{equation}
where $f_{s_x}$ and $f_{s_y}$ are the focal lengths in pixel units for the $\bm{S}_1$ and $\bm{S}_2$ directions, and  $c_{s_x}$ and $c_{s_y}$ are the optical center coordinates in pixel units for the $\bm{S}_1$ and $\bm{S}_2$ directions. The components of $\left[K_s\right]$ are set as follows:
\begin{enumerate}
	\item The optical center location $\bm{C}_s = \left(c_{s_1},\,c_{s_2}\right)^T$ is coincident to the screen image center
	\item The focal lengths $f_{s_x}$ and $f_{s_y}$ are set by noting that the smallest among the vertical and the horizontal FoVs must observe the vertical or the horizontal screen dimensions thorough the two lens system without cropping the image or observing the screen support. This implies that the screen camera must have the same FoV of the camera in the facility, but the screen resolution must be imposed ot avoid errors when displaying the image. 
\end{enumerate}  
It is worth noting that the proposed projection model from the 3D virtual world to the screen do not account for any distortion or misalignment among the facility components.\newline
Let $\mathfrak{F}$ be the facility camera projection model that maps the non-homogeneous 3D point  to the non-homogeneous distorted 2D point:
\begin{equation}
	{}_h^{\mathbb{F}}\bm{P}_d = \mathfrak{F}\left({}^{\mathcal{F}}\bm{p}\right)  
\end{equation}
where ${}_h^{\mathbb{F}}\bm{P}_d$ is the non-homogeneous distorted 2D point expressed in the screen image reference frame $\mathbb{F}$ and ${}^{\mathcal{F}}\bm{p}$ is the non-homogeneous 3D points expressed the screen camera reference frame $\mathcal{F}$.\newline
In the case of a pinhole camera model, a fundamental properties of this model is that it must represent a projection, thus it must be scale-invariant. Therefore for all scale $k$:
\begin{equation}
	 \mathfrak{F}\left({}^{\mathcal{F}}\bm{p}\right)  = \mathfrak{F}\left(k\;{}^{\mathcal{F}}\bm{p}\right)
\end{equation}
Finally, let $\mathfrak{D}$ be the error model that represents the RETINA distortions and misalignment errors. In details, $\mathfrak{D}$ maps the warping from undistorted LoSes (i.e. the LoS modeling a rendering ray) to the LoSes as observed by the facility camera when distorted by RETINA. Thus:
\begin{equation}
	{}^{\mathcal{F}}\bm{\ell}_d = \mathfrak{D}\left({}^{\mathcal{S}}\bm{\ell},\,\bm{\beta}\right)
\end{equation}
\begin{equation}
	{}^{\mathcal{S}}\bm{\ell} = \mathfrak{D}^{-1}\left({}^{\mathcal{F}}\bm{\ell}_d,\,\bm{\gamma}\right)
\end{equation}
where $\bm{\beta}$ and $\bm{\gamma}$ are sets of parameters used to model RETINA distortions and misalignment, ${}^{\mathcal{S}}\bm{\ell}$ is the undistorted LoS in the screen camera reference frame $\mathcal{S}$, and ${}^{\mathcal{F}}\bm{\ell}_d$ is the distorted LoS in the facility camera reference frame $\mathcal{F}$. Note that its is important to scale the LoSes such that ${}^{\mathcal{S}}\bm{\ell}^T\bm{s}_3 = {}^{\mathcal{F}}\bm{\ell}_d^T\bm{f}_3 = 1$.\newline
The relationship among the different distortion model are reported in Figure~\ref{fig:projectionAndDistortionRelationship} for the sake of clarity. When a screen image is displayed in RETINA, the LoSes used to generate it in the 3D virtual world are warped by the facility. These warped LoSes are then projected in the facility camera obtaining the facility image. As the stimulation in the facility implies the warping of the scene in the 3D virtual world, it is of primary importance to estimate the HIL-induced errors modeled with the mapping $\mathfrak{D}$. To do so, a detailed calibration procedure is developed as reported in Section~\ref{sec:calibration}.
Once the calibration is performed, the facility error model $\mathfrak{D}$ can be used to compensate the HIL-error at software level by artificially manipulating the image to ensure the equivalence between what is observed in the 3D virtual world and what is observed by the facility camera. In a formal, way this procedure ensures that \cite{panicucci2022tinyv3rse}:
\begin{equation}\label{eq:compensationequivalence}
	{}^{\mathcal{F}}\bm{\ell} \simeq {}^{\mathcal{S}}\bm{\ell}
\end{equation}
The compensation of HIL-induced error estimated in the calibration can be performed before and after the stimulation. If the compensation is performed before the stimulation (i.e. before displaying the image in the facility), the compensation is called upstream. Conversely, if the compensation is performed after the stimulation (i.e., after the facility camera acquisition), the compensation is called downstream. \citet{panicucci2022tinyv3rse} shows that both compensations implies Equation~\ref{eq:compensationequivalence}. The selection of the most suitable procedure between the two strongly depends on the available rendering software, the hardware-in-the-loop setup, and the simulated scenario. Moreover, for unresolved objects (e.g., stars and planets) the upstream compensation can be complemented with a subpixel correction to achieve subpixelic precision below the angular size of the screen pixel as observed by the camera. This subpixelic compensation is presented in Section~\ref{sec:subpixelcorrection} after having presented the calibration procedure in Section~\ref{sec:calibration}. In this paper, the upstream calibration is the only one employed whose functional workflow is reported in Figure~\ref{fig:upstreamCompensation}.

\subsection{Calibration Procedure}\label{sec:calibration}
\begin{figure}[!ht]
	\centering
	\begin{subfigure}{0.48\textwidth}
		\centering
		\includegraphics[width=\columnwidth]{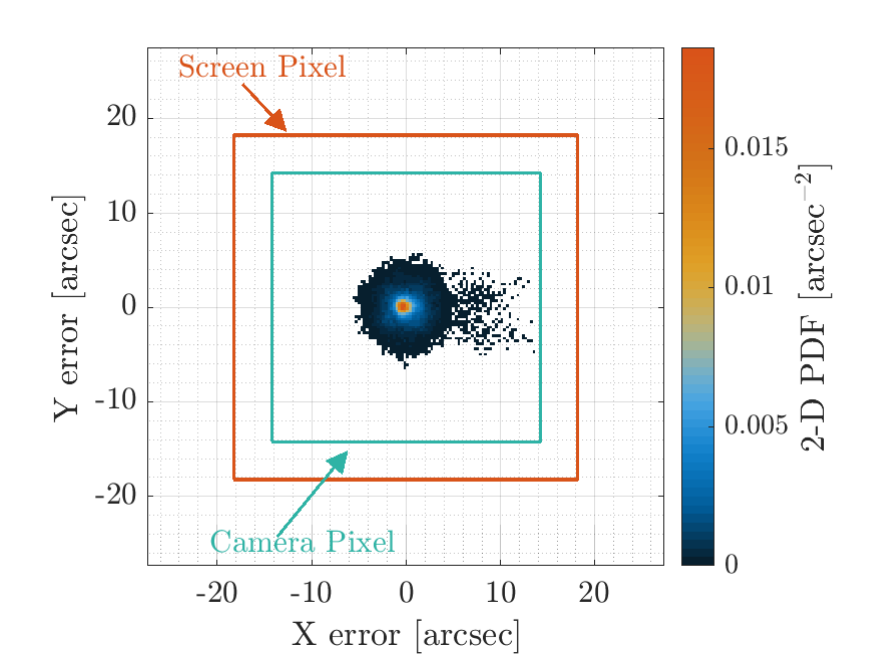}
		\caption{Probability Density Function and Cumulative Density Function of the angular error.}
		\label{fig:PDFCDFcalibration}
	\end{subfigure}
	\hfill
	\begin{subfigure}{0.48\textwidth}
		\centering
		\includegraphics[width=\columnwidth]{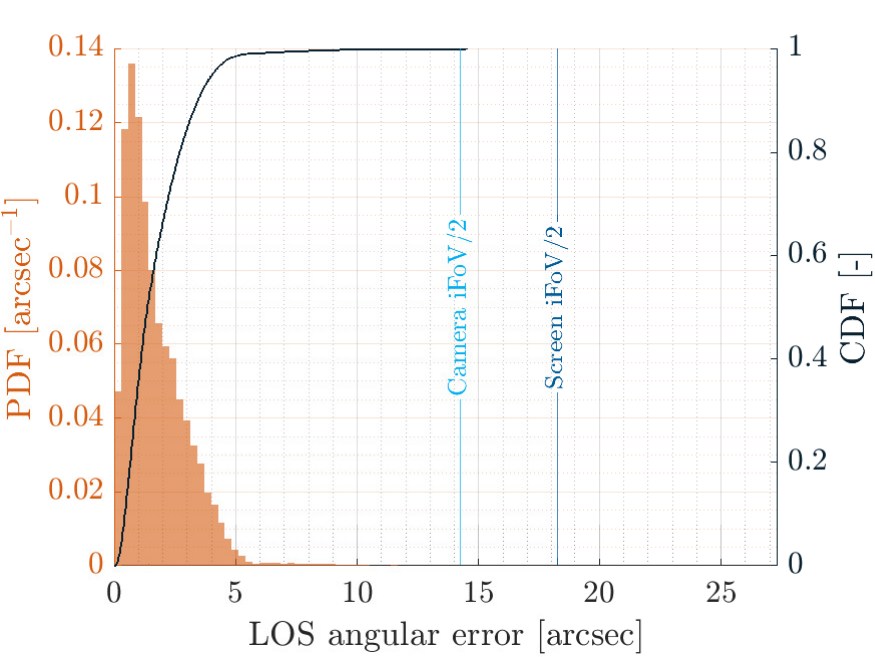}
		\caption{Two-dimensional Probability Density Function of the pixelic error.}
		\label{fig:2DPDFcalibration}
	\end{subfigure}
	\caption{Results for the calibration.}
	\label{fig:calibration}
\end{figure}

The geometric calibration is in charge of estimating the coefficients $\bm{\gamma}$ and $\bm{\beta}$ to determine the error model $\mathfrak{D}$ and its inverse. The calibration procedure can be summarized as follows:
\begin{enumerate}
	\item The facility is mounted and aligned manually in an iterative procedure that aims at centering and aligning the different components independently. This ensures that rays from the screen pass into the central part of the lenses, leading to reduced distortions and aberrations.
	\item A series of calibration patterns composed of illuminated dots are projected on the screen and a series of images are acquired by the camera mounted in RETINA.
	\item The LoS detected by the camera are compared with the ones used to generate the image displayed on the screen. By minimizing the error between the two LoS sets, $\bm{\gamma}$ and $\bm{\beta}$ are estimated.
\end{enumerate}
In more detail, the screen LoSes can be computed from the images displayed on the screen by knowing where the bright pixels are:
\begin{equation}
	{}^{\mathcal{S}}\bm{\ell} \propto [K_s]^{-1} {}^{\mathbb{S}}_h\bm{P}
\end{equation}
where $\propto$ is used as ${}^{\mathcal{S}}\bm{\ell}$ must be scaled to ensure ${}^{\mathcal{S}}\bm{\ell}^T\bm{s}_3 = 1$. Analogously, the camera LoSes are gathered from extracted centroids in facility images:
\begin{equation}
	{}^{\mathcal{F}}\bm{\ell}_d \propto \mathfrak{F}^{-1} \left({}^{\mathbb{F}}_h\bm{P}_d\right)
\end{equation}
The coefficients are estimated by minimizing the residual errors between the screen LoSes and the camera LoSes after an association step is performed. Thus:
\begin{equation}
	\hat{\bm{\beta}} = \arg\min_{\bm{\beta}}\left(\sum_{i=1}^{N_p} \left|\left| \mathfrak{D}\left({}^{\mathcal{S}}\bm{\ell},\,\bm{\beta}\right) - {}^{\mathcal{F}}\bm{\ell}_d \right|\right|\right)
\end{equation}
\begin{equation}
	\hat{\bm{\gamma}} = \arg\min_{\bm{\beta}}\left(\sum_{i=1}^{N_p} \left|\left| \mathfrak{D}^{-1}\left({}^{\mathcal{F}}\bm{\ell}_d,\,\bm{\gamma}\right) - {}^{\mathcal{S}}\bm{\ell} \right|\right|\right)
\end{equation}
where $N_p$ is the number of calibration dots in all acquired images. By using a polynomial representation \cite{panicucci2022tinyv3rse, beierle2019variable, samaan2011star}, the error model $\mathfrak{D}$ and its inverse are linear in the coefficients $\bm{\gamma}$ and $\bm{\beta}$, thus they are computed with linear least squares.\newline
For the sake of brevity, solely the results for the calibration of the inverse model error $\mathfrak{D}^{-1}\left({}^{\mathcal{F}}\bm{\ell}_d,\,\hat{\bm{\gamma}}\right)$ are reported in this document as this is the one used for the upstream compensation. Note that numerical results are gathered with the wide-FoV camera configuration, using the 25 mm objective. The calibration procedure foresees the stimulation of the camera with point-wise patterns distributed such that they cover the full camera FoV. Then the points are associated and the optimization is performed. The quality of the process is studied both on the calibration and compensation. In the first case, the estimation process is performed and the post-fit residuals are studied to understand the quality of the estimation. In the second case, the estimated parameters are used to compensate RETINA errors in never-observed point-wise patterns, and the error is directly computed between the true pattern and the observed one. This last framework is important to assess the performance of the developed method as it emulates the compensation to be performed during HIL testing.\newline
To understand the quality of the calibration and compensation process, two figures of merit are used. First, the angular error between the true screen LoS (i.e., ${}^{\mathcal{S}}\bm{\ell}$) and camera LoS undisorted exploiting the estimated coefficients (i.e., $\mathfrak{D}^{-1}\left({}^{\mathcal{F}}\bm{\ell}_d,\,\hat{\bm{\gamma}}\right)$). Second, the pixelic error between the true screen LoS projected on the facility camera (i.e., $[K_f]{}^{\mathcal{S}}\bm{\ell}$ in non-homogeneous coordinates) and the undistorted camera LoSes projected on the facility camera (i.e., $\mathfrak{F}\left(\mathfrak{D}^{-1}\left({}^{\mathcal{F}}\bm{\ell}_d,\,\hat{\bm{\gamma}}\right)\right)$ in non-homogeneous coordinates). This second figure of merit is used as it provides performance about the geometrical distribution on the image plane of the error between the two LoSes and gives insight into possible estimation biases. All results concerning calibration and compensation are obtained considering a camera with 25-mm objective, sensor size of 1/1.8 inches and F\# of 2.8. This is because this camera has the worst calibration and compensation performance for the camera envelope reported in Figure \ref{fig:feasibleCameras}, given the lower angular resolution of the pixel.\newline
As far as the calibration results are concerned, the angular post-fit residuals Probability Density Function (PDF) and Cumulative Density Function (CDF) are depicted in Figure \ref{fig:PDFCDFcalibration}. The angular error underlines that more than 99\% of the points are below 5 arcseconds during calibration, leading to the correct estimation of the distortions and misalignment among the hardware components. In the figure, the two Instantaneous FoVs (IFoVs) are reported: the camera and the screen IFoVs. Note that the screen iFoV is the screen pixel angular size as seen by the camera in the facility. The results show that the post-fit residuals are smaller than the camera IFoV, thus the errors are within the camera pixel. These are also below the screen IFoV, indicating that the fitting process is well performed. This is confirmed by the 2D PDF associated with the post-fit residual pixelic error which is reported in Figure \ref{fig:2DPDFcalibration}. The 2D pixelic error is contained within the camera pixel with a mean close to zero, indicating an unbiased solution. Only some points experience a higher error, but they are limited in number and probably at the image border. Moreover, the 2D PDF is overall consistent with a zero-mean Gaussian PDF, as expected by a least-squares approach. This analysis concludes that the facility error model parameters are well-estimated with well-distributed post-fit residuals.
\begin{figure}[!ht]
	\centering
	\begin{subfigure}{0.48\textwidth}
		\centering
		\includegraphics[width=\columnwidth]{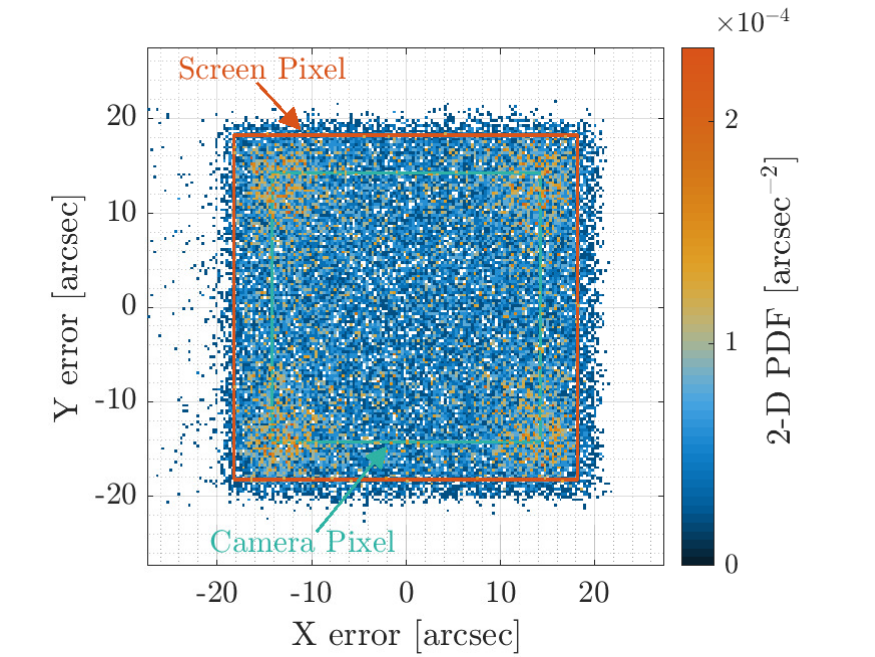}
		\caption{Probability Density Function and Cumulative Density Function of the angular error.}
		\label{fig:PDFCDFpixeliccompensation}
	\end{subfigure}
	\hfill
	\begin{subfigure}{0.48\textwidth}
		\centering
		\includegraphics[width=\columnwidth]{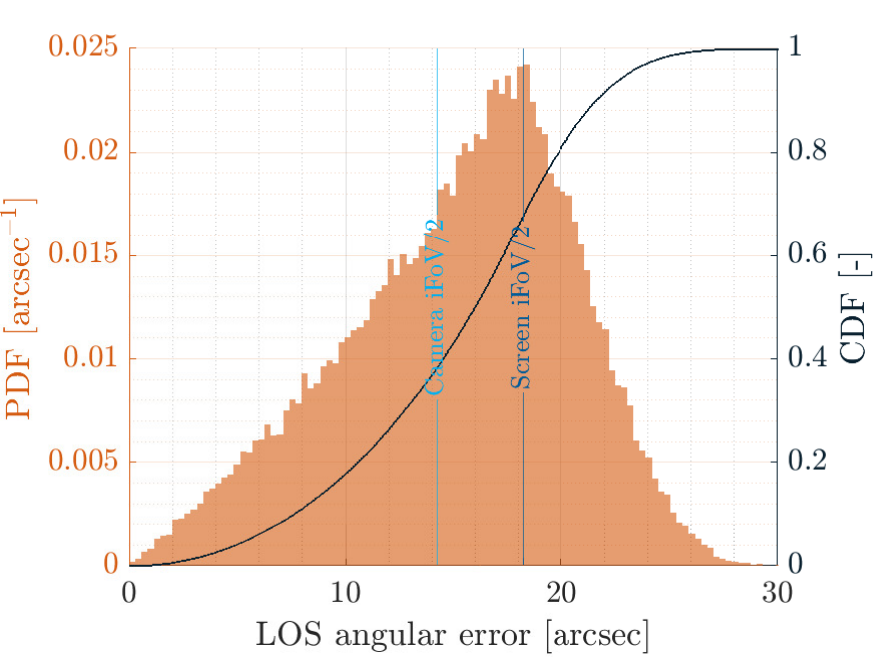}
		\caption{Two-dimensional Probability Density Function of the pixelic error.}
		\label{fig:2DPDFpixeliccompensation}
	\end{subfigure}
	\caption{Results for the compensation.}
	\label{fig:pixeliccompensation}
\end{figure}
\newline
Once the calibration is performed, the estimated parameters can be used to compensate for the HIL-induced errors. The compensation is thus tested on point-wise patterns never observed by the camera and not used for the calibration. As reported in Figure \ref{fig:upstreamCompensation}, the screen LoSes used to generate the point-wise patterns are firstly distorted exploiting the estimated RETINA error model. Then, the screen image of the pattern is generated, displayed on the screen, and acquired by the camera. Finally, the angular error and the 2D pixelic error between the original screen LoSes (i.e., before the warping procedure) and the camera LoSes are computed. The results obtained with this procedure are reported in Figures \ref{fig:PDFCDFpixeliccompensation} and \ref{fig:2DPDFpixeliccompensation} for the angular and pixelic errors, respectively. The angular error PDF and CDF show that most of the observed points are displayed with an accuracy higher than the screen IFoV. This is also consistent with the 2D PDF associated with the pixelic error. In this case, the PDF mean is close to zero, but it is not a Gaussian PDF. Indeed, as reported in \citet{panicucci2022tinyv3rse}, this PDF is consistent with a uniform PDF whose support is the screen pixel. This is compliant with the fact that the smallest portion of the screen to be illuminated is a screen pixel. Thus, when illuminating a single pixel, it is not possible to reduce the error bounds below the screen pixel boundaries. Nevertheless, this analysis shows that the upstream compensation allows stimulating the camera by reproducing the desired scene with an accuracy compliant with the smallest IFoV in the facility (i.e., the screen IFoV for the selected camera). It is worth noting that, if the screen IFoV is larger than the camera IFoV, the screen pixel would be observable in camera images, leading to a validation of the software-hardware pipeline but not to the algorithm performance (e.g., see \citet{andreis2024hardware} and \citet{regnier2024juice}). 

\subsection{Subpixel compensation}\label{sec:subpixelcorrection}

\begin{figure} 
	\centering
	\includegraphics[width=0.4\textwidth]{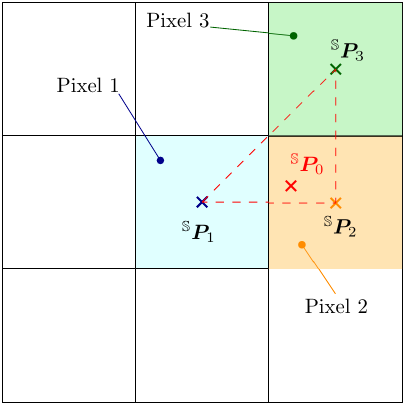}
	\caption{Example of pixel selection logic for sub-pixelic compensation correction.}
	\label{fig:subPixelCorrection_example}
\end{figure}

It is not uncommon that image processing algorithms can achieve sub-pixelic accuracy thanks to contextual information provided by neighboring pixels.
Conversely, the geometric accuracy reproduced in the facility is limited by the angular size of the screen pixel. In particular, when a single pixel is lit, a single LoS direction associated with the center of the pixel can be reproduced. This means that to perform effective simulations to validate algorithm performance, it is necessary to employ screen pixels with iFoVs comparable to the accuracy of the image processing being used. In the context of unresolved objects, these limitations are overcome by employing multiple screen pixels to achieve a more accurate geometric reproduction of the pointwise objects. Indeed, the proposed procedure cannot be applied to resolved objects as the neighboring pixels are used to stimulate the camera to reproduce the scene.\newline
The proposed approach is based on the idea that neighboring pixels to the one that is lit can be switched on as well to reduce the LoS error on the camera detector, thus increasing the accuracy. In particular, for each object, three pixels are lit. The intensity of each pixel is computed to fulfill two criteria. First, the sum of the intensities shall match the intensity of the celestial object $F_0$. Second, the coordinates of the center of mass of the reproduced shape shall coincide with the intended screen object coordinates ${^{\mathbb{S}}\bm{P}}_0$. In practice, the problem can be stated as: 

\begin{equation}\label{eq:subpixelcompensation}
	\left\{ \begin{aligned} 
		&F_0 = F_1 + F_2 + F_3 \\
		&F_0 {^{\mathbb{S}}\bm{P}}_0  = F_1 {^{\mathbb{S}}\bm{P}}_1 + F_2 {^{\mathbb{S}}\bm{P}}_2 + F_3 {^{\mathbb{S}}\bm{P}}_3
	\end{aligned} \right.
\end{equation}
where ${^{\mathbb{S}}\bm{P}}_{i}$ with $i \in \left[1, 2, 3\right]$ is the coordinates of the $i$th pixel center and $F_{i}$ with $i \in \left[1, 2, 3\right]$ is the $i$th pixel intensity.
The selection of which pixels to switch on is not trivial as it must guarantee that the problem is feasible and that solutions involve only positive values of intensities. Figure \ref{fig:subPixelCorrection_example} shows an example of the selection logic for a given configuration. In Figure \ref{fig:subPixelCorrection_example}, Pixel 1 is the one whose center is closer to  ${^{\mathbb{S}}\bm{P}}_0$. The remaining two are selected among the eight neighbors of Pixel 1. In particular, Pixels 2 and 3 are those for which the point ${^{\mathbb{S}}\bm{P}}_0$ is contained inside the triangle determined by the centers of the three pixels.  
With this selection logic, the system has always a feasible solution since the pixels are not aligned. Moreover, the solution for the three intensities has always positive values for $F_0 > 0$.
In practice, for each pointwise object, the system is solved analytically to find the correct distribution of intensities. The analytical expression is not reported here for the sake of brevity, but it can be determined from Equation \ref{eq:subpixelcompensation}.  These values are then converted into an associated digital count defined by the radiometric calibration of the facility knowing the screen calibration curve as explained in \citet{ornati2024retina} and \citet{andreis2023towards}. 

\begin{figure}
	\centering
	\begin{subfigure}{0.48\textwidth}
		\centering
		\includegraphics[width=\columnwidth]{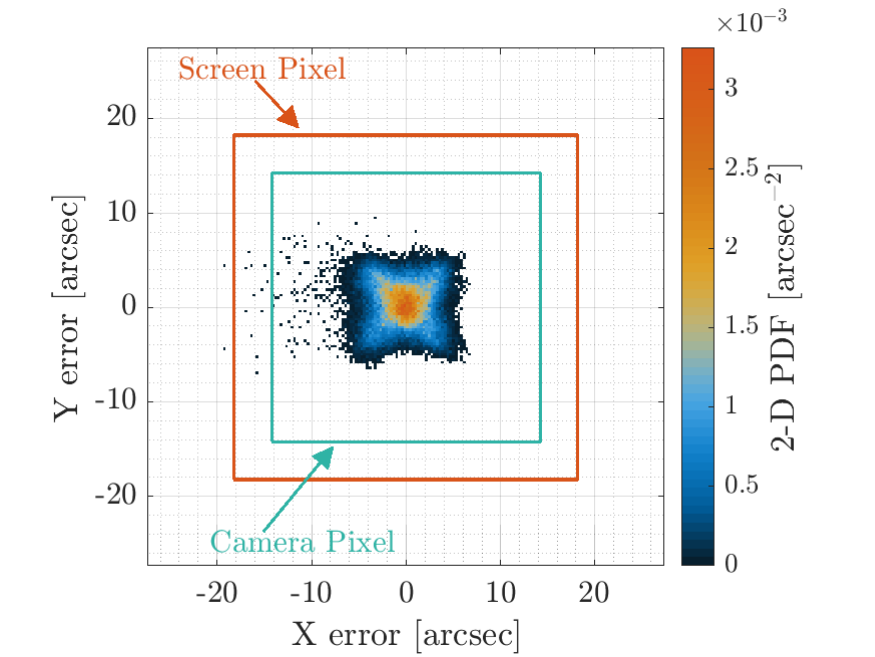}
		\caption{Probability Density Function and Cumulative Density Function of the angular error for the compensation.}
		\label{fig:PDFCDFsubpixeliccompensation}
	\end{subfigure}
	\hfill
	\begin{subfigure}{0.48\textwidth}
		\centering
		\includegraphics[width=\columnwidth]{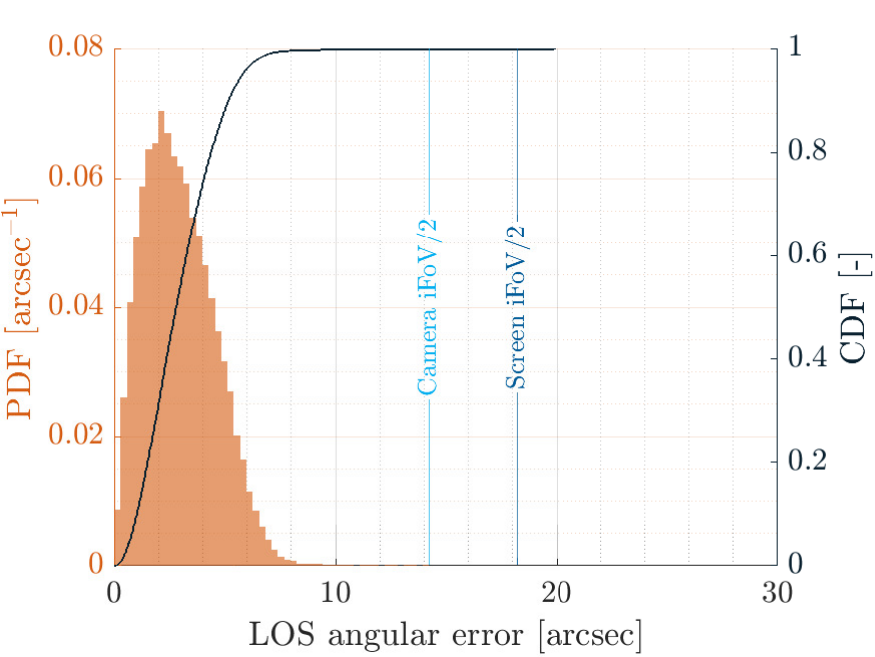}
		\caption{Two-dimensional Probability Density Function of the pixelic error for the compensation.}
		\label{fig:2DPDFsubpixeliccompensation}
	\end{subfigure}
	\caption{Results for the subpixelic compensation.}
	\label{fig:subpixeliccompensation}
\end{figure}

The effectiveness of the correction is shown in Figures \ref{fig:PDFCDFsubpixeliccompensation} and \ref{fig:2DPDFsubpixeliccompensation}, reporting the accuracy of the compensation procedure using the proposed sub-pixelic correction. In this case, the errors are considerably reduced as they are lower than 8 arcsec in 99 \% of the cases. It is worth noting that the errors almost reach the limit determined by the calibration residuals shown in Figure \ref{fig:PDFCDFcalibration} and \ref{fig:2DPDFcalibration}.

\subsection{Sensitivity to External Temperature}
\begin{figure}[ht]
	\centering
	\begin{subfigure}[t]{0.48\textwidth}
		\centering
		\includegraphics[width=0.8\columnwidth, trim={2cm 0cm 2.5cm 0cm}, clip]{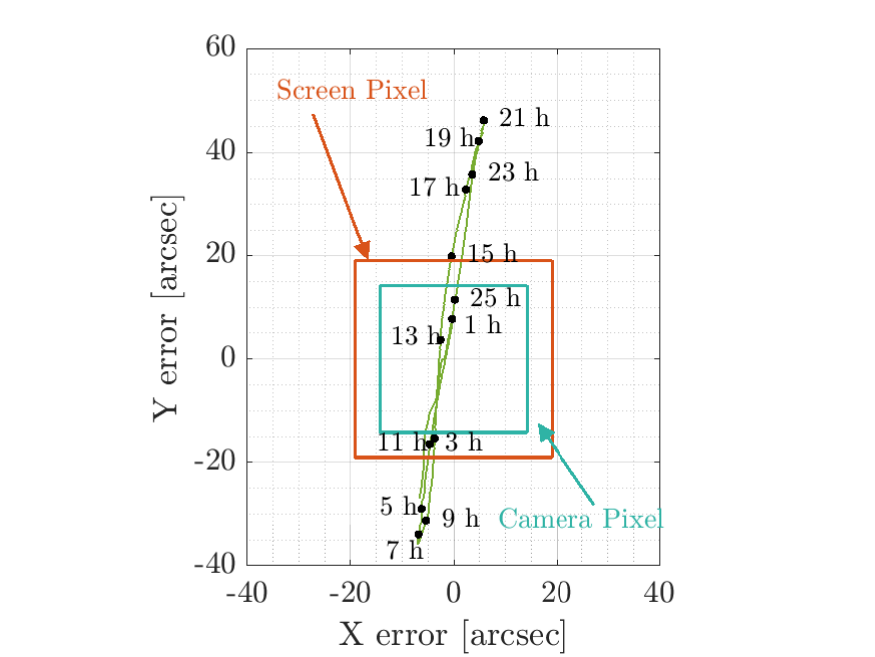}
		\caption{Evolution of the calibration accuracy in time due to thermoelastic effects.}
		\label{fig:temp_shiftXY_journal}
	\end{subfigure}
	\hfill
	\begin{subfigure}[t]{0.48\textwidth}
		\centering
		\includegraphics[width=\columnwidth]{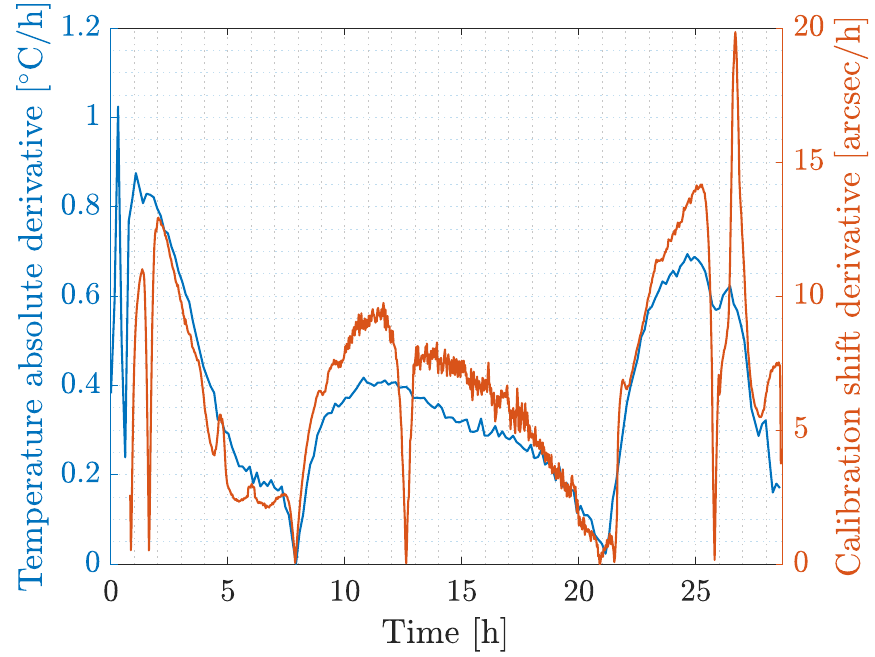}
		\caption{Correlation between the room temperature derivative and the calibration accuracy shift derivative.}
		\label{fig:temp_shift_derivative_journal}
	\end{subfigure}
	\caption{Sensitivity of the calibration to the external temperature and thermoelastic effects.}
	\label{fig:temperature_variation}
\end{figure}

As one of the purpose of RETINA is to perform  Monte Carlo campaign in HIL framework, it is of interest to characterize the facility behavior and performance in the context of external temperature variation. This is relevant as it implicitly account for thermoelastic effects in the simulation which are hard to model in rendering engine. It is worth noting that the thermostatic effects experienced in RETINA are not exactly the ones affecting the camera on orbit, but the algorithm response to the presence of thermoelastic errors is an additional evidences of the algorithm robustness when limited algorithm degradation is observed during testing.\newline
To understand the sensitivity of the estimated calibration to the temperature, a dedicated test is implemented assessing the stability of the calibration to temperature. RETINA is calibrated at a given moment and calibration parameters are estimated and stored to be used during the test. Every minute, a series of unobserved patterns are projected on the screen using the upstream compensation. Form the obtained centroids, the reprojection error is evaluated and the error mean and standard deviation for all the pattern points are computed. In the meanwhile, the temperature in RETINA is sampled via a temperature probe. Figure~\ref{fig:temp_shiftXY_journal} shows the evolution of the mean calibration offset in time during the 28 hours of the experiment. Results show that the calibration is affected by the temperature variation as the optical elements of the facility are deformed and tilted by temperature gradients. While the accuracy of the calibration is affected by these effects, it is worth noting that the standard deviation of the calibration error remains constant in time, implying a constant precision with temperature variations. Finally, to support the statement that the calibration offset is induced by the temperature variations, Figure~\ref{fig:temp_shift_derivative_journal} shows evident correlations between the temperature absolute derivative and the calibration shift derivative.\newline
In conclusion, it has been decided not to compensate this effects with a dedicated software pipeline because these are exactly the gaps that HIL simulations aim to represent when assessing the algorithm robustness and performance.

\section{Applications}\label{sec:applications}
To show the applicability of RETINA to different scenes and different hardware, two different applications are presented. First, an attitude determination algorithm exploiting stars is outlined in Section \ref{sec:attitudeDetermination}. Second, an edge detection algorithm fitting the Moon limb is expounded.\newline
On the one side, the attitude determination algorithm is presented as the observed sources (i.e., the stars) are not resolved (i.e., their apparent size is smaller than one pixel). This implies that subpixelic correction can be applied to locate accurately the star centroid in the camera images. Moreover, the camera used in this scenario uses a focal length of 25 mm with a sensor of 1/1.8 inches, leading to a diagonal full-cone FoV of 20 degrees. On the other side, the edge detection algorithm works with a resolved celestial object, thus the compensation can only achieve pixelic accuracy. Nevertheless, this application shows the versatility of the geometrical calibration procedure. Indeed, the Moon is geometrically consistent with the associated space scene, despite the calibration having been performed using point-wise calibration patterns. Moreover, the camera used has a focal length of 50 mm with a sensor of 1/1.8 inches, leading to a diagonal full-cone FoV of 10 degrees.

\subsection{Unresolved Targets: Attitude Determination}\label{sec:attitudeDetermination}
\begin{figure}[!ht]
	\centering
	\frame{\includegraphics[width=0.9\textwidth]{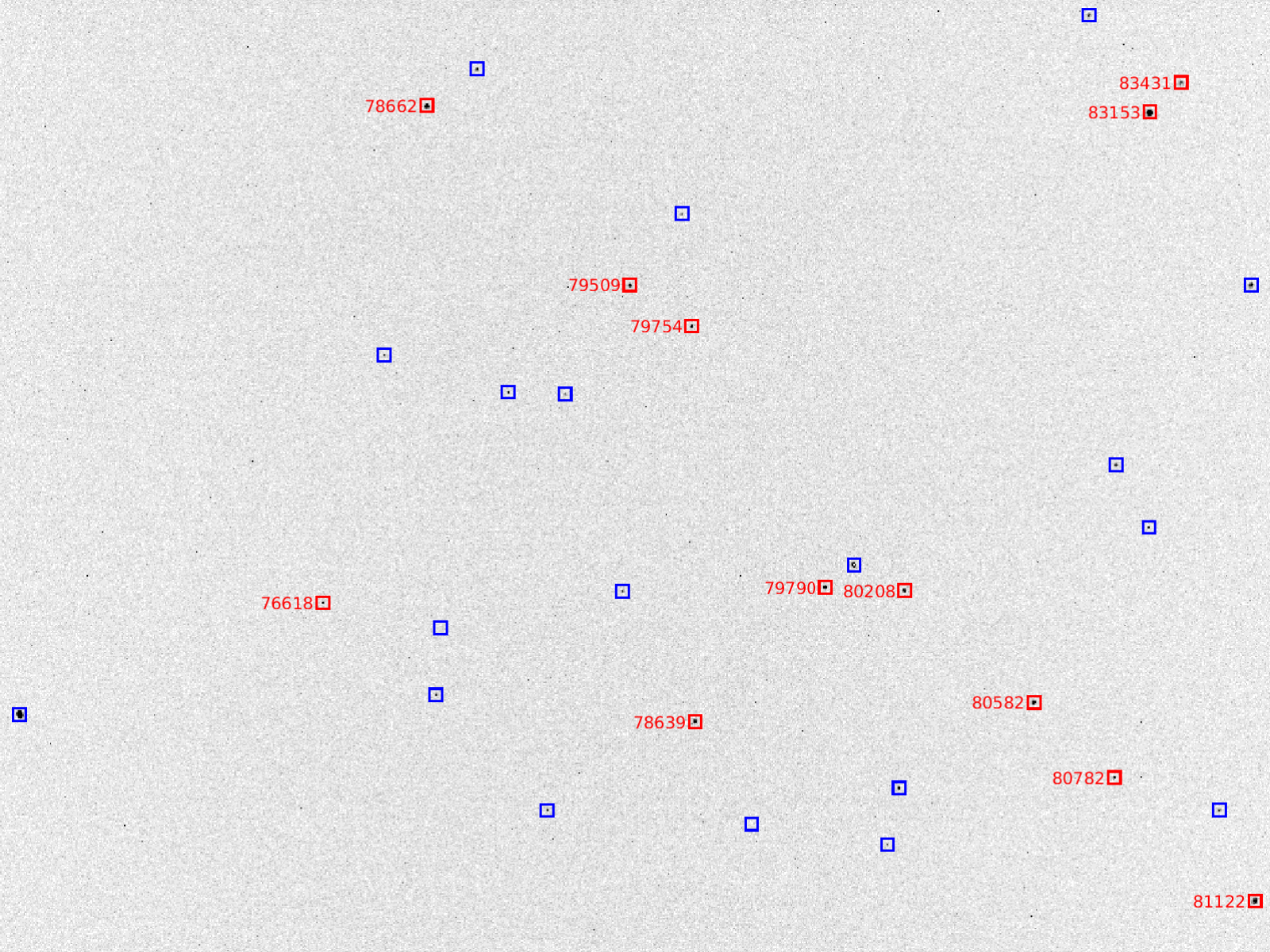}}
	\caption{Example of attitude determination on RETINA images.}
	\label{fig:imgAttituDetermination}
\end{figure}

\begin{figure}[!ht]
	\centering
	\begin{subfigure}[t]{0.48\textwidth}
		\centering
		\includegraphics[width=\columnwidth]{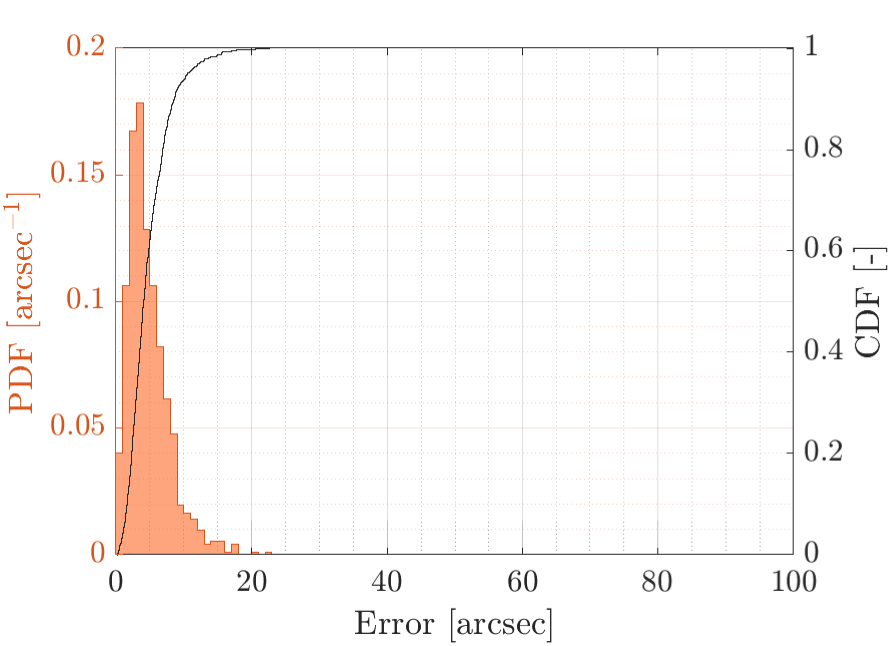}
		\caption{Cross-boresight angular error.}
		\label{fig:boresightError}
	\end{subfigure}
	\hfill
	\begin{subfigure}[t]{0.48\textwidth}
		\centering
		\includegraphics[width=\columnwidth]{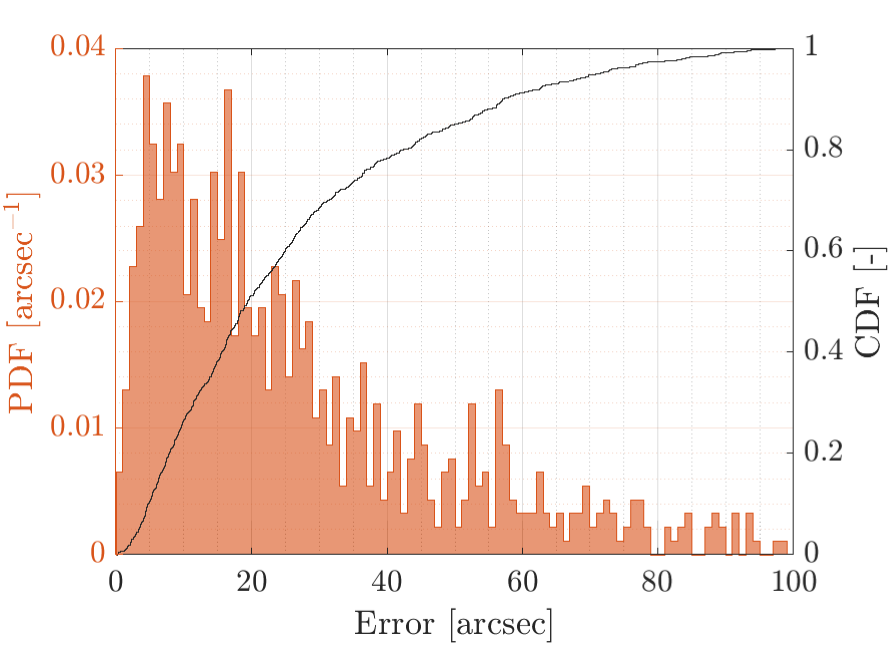}
		\caption{About-boresight angular error.}
		\label{fig:crossBoresightError}
	\end{subfigure}
	\caption{Probability Density Functions and Cumulative Density Functions of the attitude determination.}
	\label{fig:unresolvedError}
\end{figure}

The first application is the attitude determination developed in the context of the EXTREMA project \cite{di2022erc}. The proposed algorithm is a customization of the k-vector algorithm proposed by \citet{mortari2014k} bulked with the RANSAC algorithm \cite{hartley2004multiple}. The algorithm extracts bright spots in the image via a thresholding procedure and computes the associated centroids. Then, the registration between an onboard-stored star catalog is performed with the k-vector algorithm and the attitude is computed by solving the Wahba's problem. Note that the computed attitude solution robustness is increased by a consensus strategy based on RANSAC. More details about the algorithm can be found in \citet{andreis2023autonomous}. An example of the output produced by the algorithm in terms of centroid calculation and asterism identification is reported in Figure \ref{fig:imgAttituDetermination} where an image acquired in RETINA is shown in false color. In the figure, the red boxes are the identified stars with their associated star ID and the blue boxes are the centroids not associated with any star.\newline
The algorithm is tested on 1000 images generated with a dedicated rendering engine. Images are then compensated before displaying them on the screen as shown in Figure \ref{fig:upstreamCompensation}. The images are given as input to the attitude determination algorithm as taken by the camera and the algorithm output is compared against the available attitude truth.\newline
Let $\left[BN\right]$ and $[\hat{B}N]$ be respectively the rotation matrices from the inertial reference frame $\mathcal{N}$ to the real and estimated body reference frames (i.e., $\mathcal{B}$ and $\hat{\mathcal{B}}$). To compare the two attitude solutions, the cross-boresight and the about-boresight errors are compared.\newline
The cross-boresight error is computed as:
\begin{equation}
	\phi = {\rm acos}({}^{\mathcal{N}}\bm{b}_3^T{}^{\mathcal{N}}\hat{\bm{b}}_3^T)
\end{equation}
where $\bm{b}_3$ and $\hat{\bm{b}}_3$ are the third unit vector of the $\mathcal{B}$ and $\hat{\mathcal{B}}$ reference frames, respectively. Note that the third unit vector is defined as the camera boresight in this work. This error represents the angle between the estimated camera boresight direction and the real one. It is worth noting that this error is related to the pitch and yaw estimation, thus is mainly constrained by the camera and screen pixel sizes in HIL simulations \cite{rufino2001stellar, rufino2002laboratory, panicucci2022tinyv3rse}.\newline
Moreover, the about-boresight error is defined as:
\begin{equation}
	\phi = \frac{{\rm acos}({}^{\mathcal{N}}\bm{b}_1^T{}^{\mathcal{N}}\hat{\bm{b}}_1^T) + {\rm acos}({}^{\mathcal{N}}\bm{b}_2^T{}^{\mathcal{N}}\hat{\bm{b}}_2^T)}{2}
\end{equation}
where $\bm{b}_1$ and $\bm{b}_2$, $\hat{\bm{b}}_1$ and $\hat{\bm{b}}_2$ are the first and second unit vectors of the $\mathcal{B}$ and $\hat{\mathcal{B}}$ reference frames, respectively. By definition, these two unit vectors are the ones lying on the image plane. This error is computed as the mean angular error of the axes laying on the image plane, thus it is directly related to the about-boresight angular error (i.e., the spacecraft roll). It is worth noting that this angle in a HIL simulation is not constrained only by camera and screen resolution, but also by the star asterisms recognized by the attitude determination algorithm \cite{rufino2001stellar, rufino2002laboratory, panicucci2022tinyv3rse}.\newline
The attitude is estimated in 97.2\% of the cases which is consistent with software-in-the-loop simulations \cite{andreis2024hardware}. Among the remaining samples (i.e., 972), 1.03\% has an attitude Absolute Performance Error (APE) greater than 300 arcseconds, which is considered the limit for precise attitude determination.  Moreover, Figures \ref{fig:boresightError} and \ref{fig:crossBoresightError} show the cross-boresight and about-boresight errors as computed from the 1000 images acquired in RETINA. As expected, the cross-boresight angular solution is estimated with greater precision than the about-boresight one. It is worth noting that cross-boresight angular error is of the order of magnitude of the calibration error (see Figure \ref{fig:PDFCDFsubpixeliccompensation} and \ref{fig:2DPDFsubpixeliccompensation}). Indeed, the main contributions to the cross-boresight angular solution are:
\begin{enumerate}
	\item The precision in detecting the real centroids in camera images that is usually around 0.3 pixels.
	\item The precision in displaying a single star on the screen using the subpixelic correction reported in Section \ref{sec:subpixelcorrection}. 
\end{enumerate}
On the contrary, the about-boresight angular solution is also influenced by the position of the identified stars in the image plane. Indeed, the closer the start to the camera cross-boresight, the higher the attitude error is. It is worth noting that in HIL optical facilities the stimulation is usually less accurate at the external part of the camera FoV (see Figure \ref{fig:finalDesignSpotDiagram25mm}, Figure \ref{fig:finalDesignSpotDiagram50mm} and Table \ref{tab:finalDesignSpotDiagramTable}). This is mainly because light passes in the external part of the lenses and it is keen to be more distorted and aberrated. Nevertheless, the about-boresight angular solution obtained in HIL simulations is consistent with state-of-the-art attitude determination algorithm performance thanks to the optimized optical design allowing precise stimulation of the camera.

\subsection{Resolved Targets: Edge Detection}\label{sec:edgedetection}
\begin{figure}[!ht]
	\centering
	\resizebox{\textwidth}{!}{%
		\begin{tikzpicture}[spy using outlines={rectangle, width=6.5cm, height=6.5cm, magnification=3, blue!40, connect spies}]
			\node[ inner sep=0pt] (image) at (2,1)
			{\frame{\includegraphics[width=0.8\textwidth]{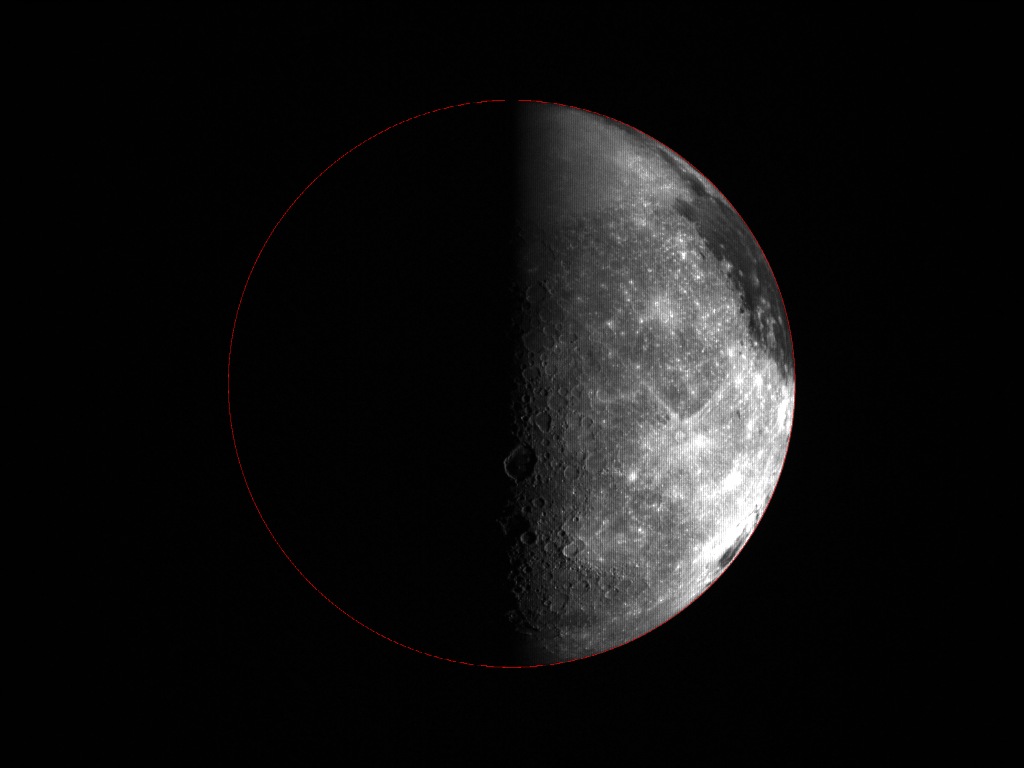}}};	
			\begin{axis}[xmin=-1,xmax=1,ymin=-1,ymax=1,hide axis]
				\coordinate (spy point) at (axis cs:0.55,-0.3);
				\coordinate (magnifying glass) at (rel axis cs: -0.78,0.8);
			\end{axis}
			\spy [white] on (spy point) in node at (magnifying glass);
	\end{tikzpicture}}
	\centering
	\caption{An example of Moon image acquired in RETINA with the theoretical Moon limb reported in red to show the geometrical consistency of the calibration.}
	\label{fig:imgMoonValidation}
\end{figure}

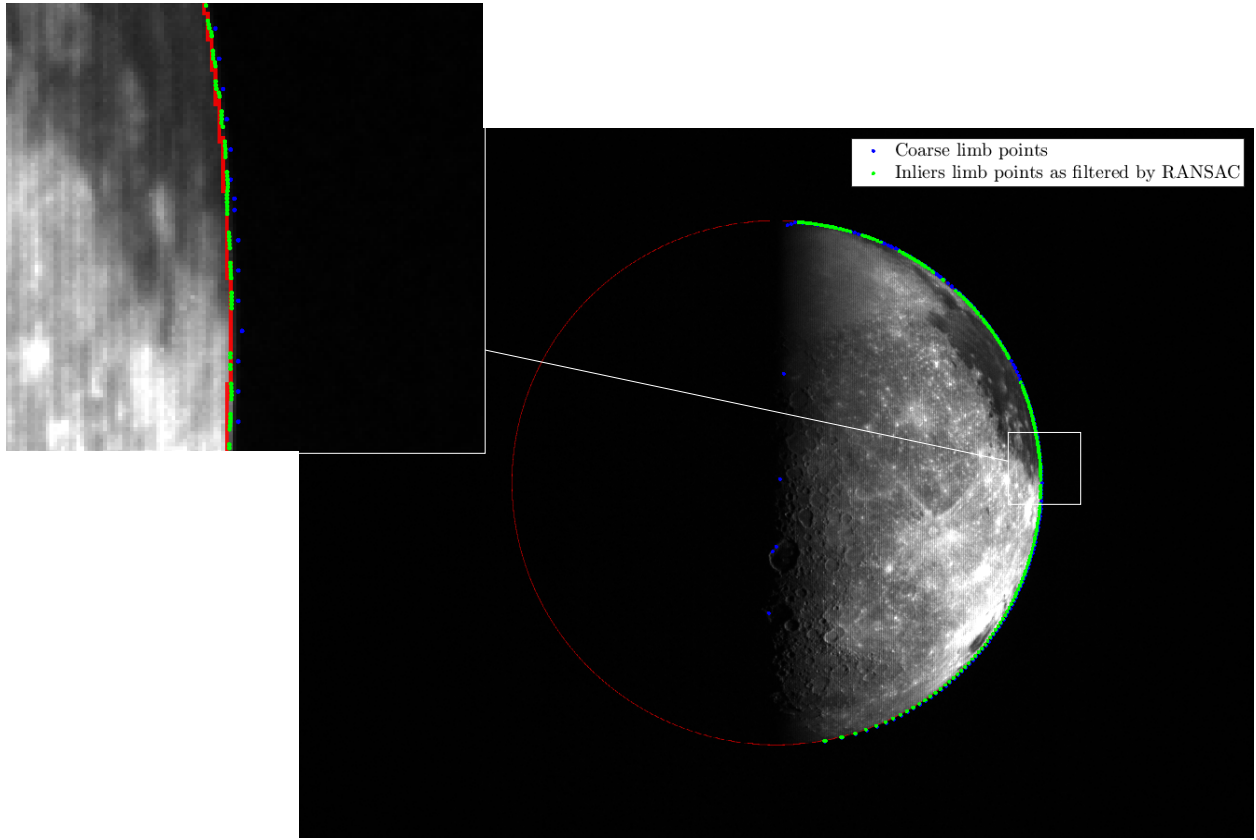
\begin{figure}[!ht]
	\centering
	\resizebox{\textwidth}{!}{%
		\begin{tikzpicture}[spy using outlines={rectangle, width=6.5cm, height=6.5cm, magnification=10, blue!40, connect spies}]
			\node[ inner sep=0pt] (image) at (2,1)
			{\includegraphics[trim={0cm 0cm 0cm 0cm}, clip, width=0.8\textwidth]{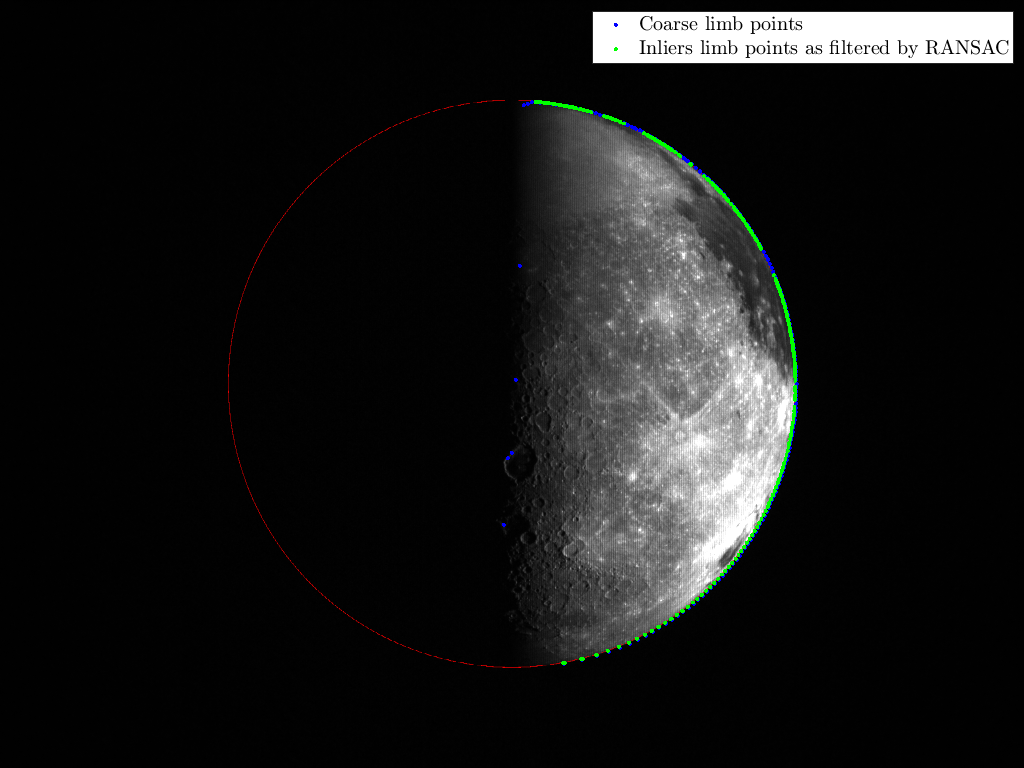}};	
			\begin{axis}[xmin=-1,xmax=1,ymin=-1,ymax=1,hide axis]
				\coordinate (spy point) at (axis cs:0.65,-0.5);
				\coordinate (magnifying glass) at (rel axis cs: -0.78, 0.8);
			\end{axis}
			\node[ inner sep=1pt, rectangle, draw=white] (imageZoom) at (rel axis cs: -0.78,0.8)
			{\includegraphics[trim={2.8cm 0cm 2.8cm 0cm}, clip, width=0.4\textwidth]{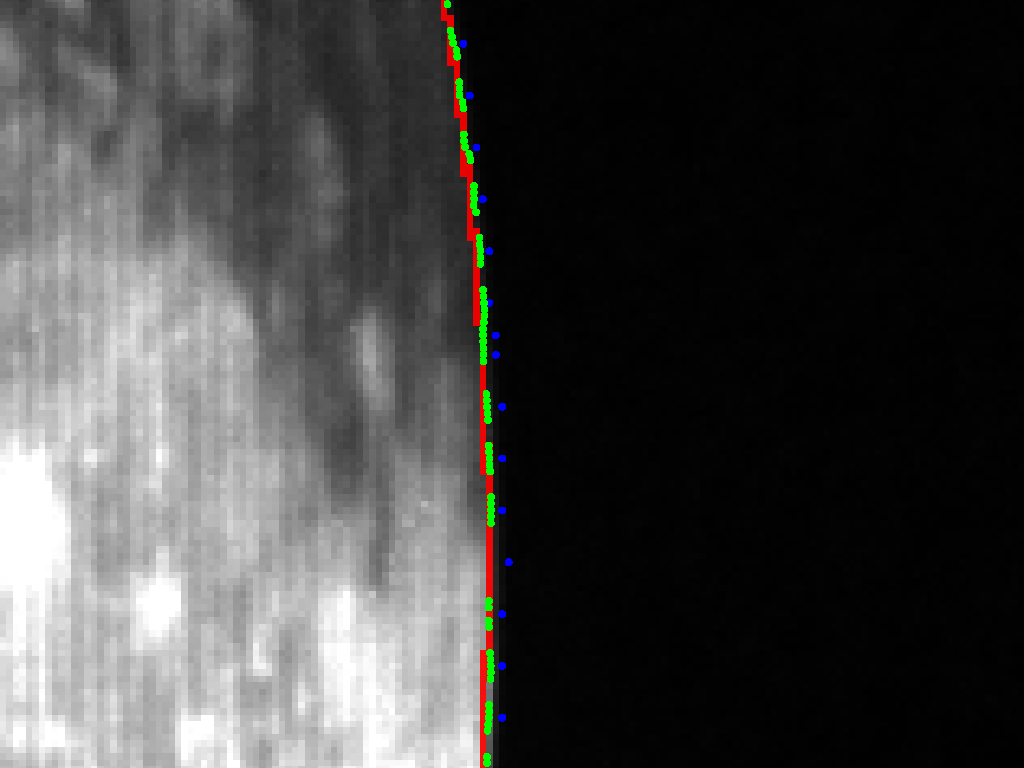}};
			\draw [draw=white] (5.2,1.7) rectangle (6.2,0.7);
			\draw [draw=white] (5.2,1.3) -- (rel axis cs: -0.295,0.5);
	\end{tikzpicture}}
	\centering
	\caption{An example of Moon image acquired in RETINA reporting the points detected on the edge and the theoretical Moon limb.}
	\label{fig:imgMoonResults}
\end{figure}

The second application relates with the detection of points belonging to the Moon limb from images developed in the context of LUMIO projects \cite{panicucci2024vision}. This application is of interest for two main reasons for RETINA. First, it enable the facility validation with resolved objects where the upstream compensation cannot be complemented with subpixelic correction. Therefore, the accuracy of the celestial object features as observed by the camera is the one reported in Figures \ref{fig:PDFCDFpixeliccompensation} and \ref{fig:2DPDFpixeliccompensation}. In this context, it is worth noting that the pixel of the screen as seen from the camera is larger than the camera pixel (see Figures \ref{fig:PDFCDFpixeliccompensation} and \ref{fig:2DPDFpixeliccompensation}), leading to less accurate Moon limb stimulation at camera level. Second, the geometrical calibration estimate the distortion coefficient switching on single pixels - thus, unresolved celestial objects - on the screen. The correct stimulation of the camera with a resolved celestial object is an independent validation of the geometrical calibration procedure and a proof of its versatility. An example of the quality of the obtained stimulation is reported in Figure \ref{fig:imgMoonValidation} where an image acquired in RETINA is shown along with the theoretical projection of the ellipsoid approximating the Moon. Figure \ref{fig:imgMoonValidation} reports also a zoom on a portion of the edge to underline the accuracy obtained in the stimulation.\newline
The proposed algorithm is inspired by the precursor work of \citet{christian2017accurate} and it is composed of several steps:
\begin{enumerate}
	\item The image is scanned to identify a coarse estimation of points laying on the Moon edge. The proposed scanning technique has been implemented as reported in \citet{panicucci2024vision} to speed up the algorithm an avoid expensive calculations on the images at the cost of increasing the number of outliers. 
	\item Patches of 7x7 pixels are extracted around the coarse edge location and a first solution for the edge is found using the Laplacian of Gaussian kernel. Then the solution is refined at subpixelic accuracy with the Zernike moments method outlined in \citet{christian2017accurate}.
	\item Outliers are rejected using the RANSAC algorithm \cite{hartley2004multiple} fitting a circle on the found subpixelic edge points.
\end{enumerate}
This algorithm is tested on more than 4000 images of the first Nav\&Eng cycle of the LUMIO trajectory \cite{panicucci2024vision}. The selected hardware is chosen to be as similar as possible to the LUMIO-Cam to increase the scenario representativeness in view of increasing the VBN algorithm TRL.
Images are rendered offline and then compensated with the pixelic upstream procedure. The acquired images are then given to the edge determination algorithm which provide three different sets of points: the coarse edge location, the subpixelic edge location, and the inliers identified by the RANSAC. To compute the accuracy of the point set the radial error $\epsilon_{r}$ has been computed for each point with the following formula:
\begin{equation}
	\epsilon_{r} = \left|\left|\bm{L}_{\rm edge} - \bm{C}_{\rm Moon}\right|\right| - r_{\rm Moon}
\end{equation}
where $\bm{L}_{\rm edge}$ is the edge point location as estimated by the algorithm, while $\bm{C}_{\rm Moon}$ and $r_{\rm Moon}$ are the true center and the true radius of the circle approximating the Moon projection.\newline
An example of the edge detection output is reported in Figure \ref{fig:imgMoonResults}. The Moon projection is reported in red, while the coarse limb points and the RANSAC inliers points are in blue and green, respectively. It is orth noting that some points, due to the scanning procedure, are also extracted on the terminator line, but are then rejected from the RANSAC algorithm. Moreover, it is possible to note that the RANSAC inliers points are qualitatively closer than the coarse edge location to the Moon projection.
\begin{figure}[!ht]
	\centering
	\begin{subfigure}[t]{0.48\textwidth}
		\centering
		\includegraphics[width=\columnwidth]{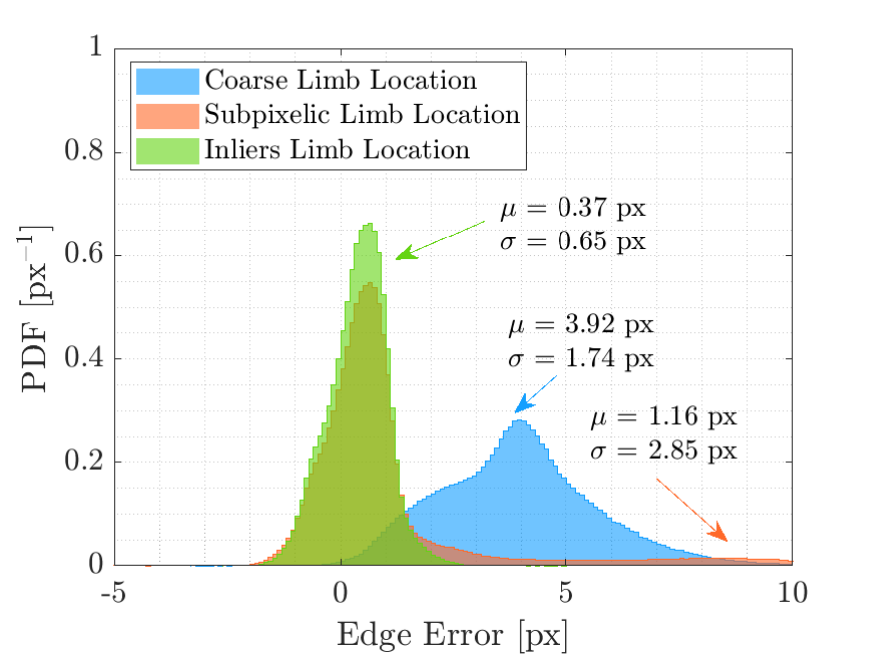}
		\caption{Probability Density Functions.}
		\label{fig:pixelErrorPDF}
	\end{subfigure}
	\hfill
	\begin{subfigure}[t]{0.48\textwidth}
		\centering
		\includegraphics[width=\columnwidth]{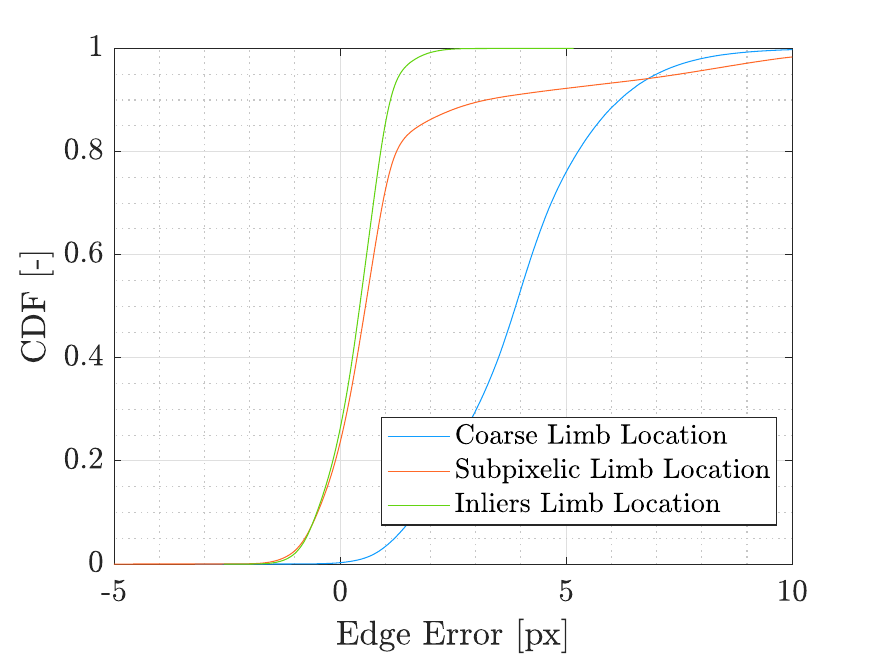}
		\caption{Cumulative Density Functions}
		\label{fig:pixelErrorCDF}
	\end{subfigure}
	\caption{Statistics of the edge detection steps for the different points extracted from the image.}
	\label{fig:pixelError}
\end{figure}
This is also confirmed by the statistics of the radial error computed for all the three points sets extracted from all the images of the LUMIO dataset. The PDFs and the CDFs for the three points sets are reported in Figures \ref{fig:pixelErrorPDF} and \ref{fig:pixelErrorCDF} respectively. It is worth noting that the refinement procedure work as expected also in on HIL simulations, increasing the accuracy of the edge point location in the different algorithmic steps. Indeed, the coarse edge points are the most biased ones with a very high standard deviation. Then the subpixelic correction step correct the bias for the majority of the points, except for some outliers that are rejected effectivelly by the RANSAC. Indeed, the RANSAC inliers are an accurate solution of the limb location in the image with a reduced standard deviation and bias. Edge location performance are slightly worse than the simulations performed with synthetic images reported in \citet{panicucci2024vision}, but this is mainly due to the fact that the camera pixel is smaller than the screen pixel as observed by the camera. 

\section{Conclusion}\label{sec:conclusion}
This work presents the design, developments, performance and applicability of the hardware-in-the-loop optical facility RETINA. The facility is designed to have variable magnification, enabling the use of cameras with different field of views, sensor sizes, and focal length. The design drivers are analyzed taking into account the allocation of the components in the facility and the vignetting due to physical size of the lenses. The analysis leads to the selection of the lens focal lengths to be used in the facility. A detailed analysis in a state-of-the-art optical design software underlines the presence of high distortions and aberrations, compromising the camera stimulation and the possibility of testing hardware and algorithm in the facility. Therefore, a dedicated low-aberration design is performed using multiple lenses that reduces the distortions and aberrations in the facility in the desired operational range. A geometrical calibration algorithm is also discussed that enable the estimation of RETINA misalignment and centering errors. These estimated parameters are then used at software level to compensate these effect and obtain an accurate stimulation of the hardware located in RETINA. The compensation performance shows that a single point can be correctly observed by the camera with an accuracy of less than 30 arcseconds for a pixelic compensation and less than 10 arcseconds for a subpixelic one. Example of the use of RETINA are also reported for resolved and unresolved celestial objects. First, an attitude determination algorithm is tested to show the performance of the facility with wide field of views, unresolved objects, and subpixelic compensation. Then, an edge detection algorithm is studied to underline the applicability of RETINA with narrow field of views, resolved objects, and pixelic compensation. Both application show that RETINA is a versatile and accurate hardware-in-the-loop optical facility that can be used to test VBN and IP algorithm with different scenario, hardware, and celestial objects. 

\section*{Acknowledgments}
This research is part of EXTREMA, a project that has received funding from the European Research Council (ERC) under the European Union’s Horizon 2020 research and innovation programme (Grant Agreement No.\ 864697). The authors would like to thank Dr. Eleonora Andreis for the algorithm reported in Section \ref{sec:attitudeDetermination}.

\bibliographystyle{unsrtnat}
\bibliography{sample.bib}

\end{document}